# Observations of high definition symmetric quasi-periodic oscillations in the mid-latitude ionosphere with LOFAR.

H. Trigg[1], G. Dorrian[1], B. Boyde[1], A. Wood[1], R. A. Fallows[2], M. Mevius[3]

[1]SERENE Group, Electronic, Electrical and Systems Engineering, University of Birmingham, UK

[2]RAL Space, United Kingdom Research and Innovation, Science & Technology Facilities Council, Rutherford Appleton Laboratory, Harwell Campus, Oxfordshire, UK

[3]ASTRON – The Netherlands Institute for Radio Astronomy, Oude Hoogeveensedijk 4, 7991 PD Dwingeloo, The Netherlands

## Abstract

We present broadband ionospheric scintillation observations of highly defined symmetric quasi-periodic oscillations (QPO: Maruyama 1991) caused by plasma structures in the mid-latitude ionosphere using the LOw Frequency ARray (LOFAR: van Haarlem et al., 2013). Two case studies are shown, one from $15^{th}$. December 2016, and one from $30^{th}$. January 2018, in which well-defined main signal fades and secondary diffraction fringing are observed. In particular, the broadband observing capabilities of LOFAR permit us to see considerable frequency dependent behaviour in the QPOs which, to our knowledge, is a new result. We extract some of the clearest examples of scintillation arcs reported in an ionospheric context, from delay-Doppler spectral analysis of these two events. These arcs permit the extraction of propagation velocities for the plasma structures causing the QPOs ranging from 50 – 200 ms$^{-1}$, depending on the assumed altitude. The spacing between the individual plasma structures ranges between 5-20 km. The periodicities of the main signal fades in each event and, in the case of the 2018 data, co-temporal ionosonde data, suggest the propagation of the plasma structures causing the QPOs is in the E-region. Each of the two events is accurately reproduced using a Gaussian perturbation phase screen model. Individual signal fades and enhancements were modelled using small variations in total electron content (TEC) amplitudes of order1 mTECu, demonstrating the sensitivity of LOFAR to very small fluctuations in ionospheric plasma density. To our knowledge these results are among the most detailed observations and modelling of QPOs in the literature.

# 1. Introduction

Ionospheric quasi-periodic oscillations (QPO) are characterised by wave-like recurrent variations in the received power of trans-ionospheric or backscattered radio signals. In the mid-latitude ionosphere they have been attributed to filamentary plasma structures extended along inclined field lines in sporadic E-layers (Woodman et al., 1991; Saito et al., 2006). The filaments are extended in altitude over a larger range than typically associated with normal sporadic-E layers and are associated with the appearance of spread-E in ionograms (Riggin et al., 1986; Barnes, 1992). VHF radar observations have established altitudes for these structures as high as 130 km, shown that they vary in altitude by up to 30 km as they propagated, and have periodicities comparable to the local Brunt-Väisälä frequency (From & Whitehead, 1986; Yamamoto et al., 1992; Hysell et al., 2014).

Yamamoto et al., (1991) categorised their occurrence as 'quasi-periodic' or 'continuous', with the continuous category being typically observed post-sunrise and manifesting as a series of repeated echo features which persist over several hours. These features were observed to propagate at ~120 ms$^{-1}$. The quasi-periodic category were, again, caused by plasma structures that were extended in altitude and approximately field-aligned, however only a few of them, or even just one, might be seen in a given observation. Maruyama et al., (2000), modelled altitude-extended complex sporadic-E structures seen in VHF radar echoes as a two-layer E-layer model. The principle argument being that the radio signal variations are a convolution of normal background night time E-layer with a second E-layer, separated in altitude, and containing a relatively dense plasma cloud with high electron density. The plasma cloud had a linear shape and was of approximately 1-km in transverse scale size.

The general form of individual QPOs were categorised into two broad categories by Maruyama (1991; 1995). Type 1 are asymmetric, being characterised by a rapid increase or decrease in received signal power which is then followed by a series of weaker signal oscillations akin to the damped oscillations of a bell after it has been struck. Type 2 are symmetric, in which ringing bell-like signal oscillations both preceded and followed a large signal fade. Indeed, earlier work by Doan & Forsyth (1978) referred to QPO secondary signal oscillations as 'ringing irregularities'. These secondary signal oscillation features have been attributed to Fresnel diffraction patterns being produced when the radio signal passes through a region of steep plasma density gradient (Bowman, 1989; Maruyama, 1995); the appearance of symmetry or asymmetry thus being an indicator as to whether the causal plasma structures themselves have a symmetric or non-symmetric form as they propagate with differing plasma density gradients on the leading or trailing edges.

Any exo-atmospheric radio source observed from the ground will present a signal that is a convolution of itself and variations imposed upon it during passage through the Earth's ionosphere, usually termed 'ionospheric scintillation'. If the source is natural and at astrophysical distances, such as a pulsar or radio galaxy, then scintillation may also be applied to the signal from passage through the interstellar medium and the solar wind. This paper focusses on examples of ionospheric scintillation and, while separating the contributions of these various signatures is not our focus, the interested reader is directed to Fallows et al., (2016). Broadly speaking, ionospheric scintillation observations can be separated from solar wind scintillation on the basis of plasma structure scale size, propagation velocity and direction, and plane-of-sky distance between the Sun and the raypath. Solar wind plasma irregularities detected by interplanetary scintillation can be of order 100 000 km in scale size or larger (e.g. Dorrian et al., 2010), propagate at minimum speeds in excess of 300 kms$^{-1}$ (Breen et al., 1996), and follow an approximately radial outflow direction from the Sun.

The international LOw Frequency ARray (LOFAR: van Haarlem et al., 2013) is an integrated network of ground based broadband radio telescopes operating ostensibly between 10-250 MHz. At the time of writing LOFAR consists of 52 ground stations, most of which are located in the Netherlands, however there are also 14 international stations in countries across Europe including, the Republic of Ireland, the UK, Germany, France, Latvia, Sweden, and Poland. In the Netherlands there is a dense cluster of 24 stations geographically co-located which are referred to as the LOFAR 'core' stations and given the station ID prefix of CSXXX. Further afield but still within the Netherlands are more stations which are referred to as 'remote' stations and given station ID prefixes of RSXXX.

Each LOFAR station consists of two clusters of antennas, the Low Band Antennas (LBA) which operate from 10-90 MHz, and the high-band antennas (HBA) which operate from 110-250 MHz. Frequencies at the extrema of the nominal LBA and HBA bandwidths are often filtered due to heavy RFI, however RFI may still be encountered at other frequencies (e.g. Vruno et al., 2023). In this paper we use data from the LBAs in the UK, Dutch, and Polish stations. The LBAs at each station are a cluster of 96 dual-polarisation crossed-dipole antennas; here we use data from the LBAs recorded between 21.8 – 76.1 MHz in 100 channels, with channel spacing optimised for even sampling of wavelength.

As most natural radio sources are broadband emitters and LOFAR is a broadband receiver, we are able to observe scintillation across many frequencies simultaneously. Furthermore, the geographical distribution of LOFAR ground stations throughout Europe lends itself well to tracking the propagation and evolution of ionospheric features in the LOFAR field-of-view. Numerous recent studies have taken advantage of these characteristics to study the mid-latitude ionosphere at relatively unprecedented levels of detail (e.g. Mevius et al., 2016; de Gasperin et al., 2018; Fallows et al., 2020; Boyde et al., 2022, Dorrian et al., 2023).

Ionospheric variations on LOFAR radio signals are generally classed as 'diffractive scintillation' if they are generated by plasma structures in the ionosphere that are smaller than the local Fresnel scale (e.g. Fallows et al., 2020). They are characterised by rapid and essentially randomised variations in amplitude, usually across all frequencies, with individual scintilla typically having lifetimes of order ~10 seconds. Structures in the ionosphere which are larger than the local Fresnel scale appear with a more periodic and longer lasting signature of up to several minutes (e.g. Boyde et al., 2022). In these cases the ionosphere is behaving more like a large concave lens which undergoes steady deformation as the plasma structure moves through the raypath; the signatures generated in these instances are therefore a consequence of lens-like refraction.

In this paper we present two case studies of LOFAR observations of type 2 QPOs, one in a post-sunset context from 15$^{th}$. December 2016, and another in a post-sunrise context from 30$^{th}$. January 2018, and thus both from approximately solar minimum. Geophysical conditions in both cases were very quiet, with Kp index not exceeding 2+, and F10.7 solar radio flux not exceeding 70. The high frequency resolution and high time resolution capabilities of LOFAR permit a hitherto unprecedented level of detail to be extracted from these phenomena. The first case, from 2016 we refer to as a continuous-type oscillation as we observe 15 individual oscillation events over at least 30-minutes (the events overlap with the beginning and end of the observing window), with each one having similar characteristics. The second case, from 2018, is seen in full from start to finish and is more quasi-periodic in nature, with 6 individual oscillation events over an approximately 10-minute interval but with particularly clear ringing irregularities visible in each event.

The observations are presented in section 2; in section 3 we perform some more detailed analysis which takes advantage of the unique broadband observing characteristics of LOFAR which has not been possible with previous observations of QPOs. In section 4 a brief overview of the model is presented (full details can be found in Boyde et al., 2022), with results from the QPO modelling. To our knowledge the level of detail and the frequency dependent behaviour of QPOs seen here is unprecedented in the literature and represents the first broadband ionospheric scintillation observations of type 2 QPOs.

## 2. LOFAR Observations

The data used for this study is from LOFAR observations made of radio sources Cassiopeia A (Right Ascension: 19h59m28s, Declination: 40.73°) and Cygnus A (Right Ascension: 23h23m24s, Declination 58.82°) on two separate occasions: 30 January 2018 between 04:22 and 05:00 UT, and 15 December 2016 between 18:17 and 19:50 UT.

Samples are taken at a time resolution of 0.01049 seconds, and the frequency band covers 100 frequency channels between 21.8 and 76.1 MHz which were approximately evenly distributed by wavelength. LOFAR station beam sensitivity reduces at lower elevations, and the elevation changes appreciably throughout the observing window. Consequently, to excise RFI, each individual frequency channel was median filtered using a sliding window of 50 data points. Any median filtered data points which exhibited a signal power exceeding 5σ above the standard deviation for that channel were removed. Channels still contaminated heavily with RFI after this process were fully removed and are visible as horizontal white lines in the dynamic spectra. A polynomial de-trending function was then subtracted from each frequency channel to account for elevation dependencies. Actual data use was restricted to channels between 22.5 and 64.8 MHz even though data from a slightly wider range was available, as frequencies at the extrema of the available bandwidth were heavily contaminated by RFI. To ease the computational burden of RFI mitigation and elevation dependence on the remaining channels, the data have also been down-sampled in time by a factor of 30, giving a time resolution of ~0.3 seconds.

### 2.1. January 2018

Figure 1 shows the dynamic spectra of data from LOFAR station PL612, located in Poland (latitude: 53.6°, longitude: 20.6°), from the 2018 observation of Cygnus A. Cygnus A was at an azimuth and elevation of 66.77° and 36.83° respectively at the beginning of the observation, increasing slightly to 70.64° and 38.30°, respectively, by 05:00 UT. The dynamic spectra cover the first 1590 seconds of the Cygnus A observation sliced into 3 sections of 530 seconds each. A series of repeating v-shaped signal fades bounded by symmetric signal enhancements are clearly visible and extend from the beginning of the observation at 04:22

UT to at least 04:50 UT with, possibly, fainter examples later on. The features are broader in time at lower frequencies. A key characteristic of the features is their high periodicity, with 15 individual events identified in the first 1600 seconds. As the phenomenon was ongoing in the field-of-view of the LOFAR station at the start of the observing window its full extent in time is unknown.

Given the commonality of form shared across all features they have been labelled with numbered feature 4 as a reference, as 'A', 'B', and 'C'. In the forthcoming text the parts of the features labelled 'A' are referred to as 'ringing irregularities' or 'secondary fringing', the parts labelled 'B' as 'boundary signal enhancements', and 'C' as the 'v-shaped' or 'main' signal fades. The green lines denote the start/end point of each repetition of the periodic feature and are located at the point in the secondary fringing region at which the direction of the frequency-dependent curvature changes from right to left. These positions have been approximated by eye, excluding the region between 600 and 655 seconds in which no suitable fringe boundary could be identified. The only exception being the transition from events 5 to 6 in which the secondary fringing region is less well-defined, possibly due to the presence of a fainter and less distinct v-shaped fade at ~630 seconds.

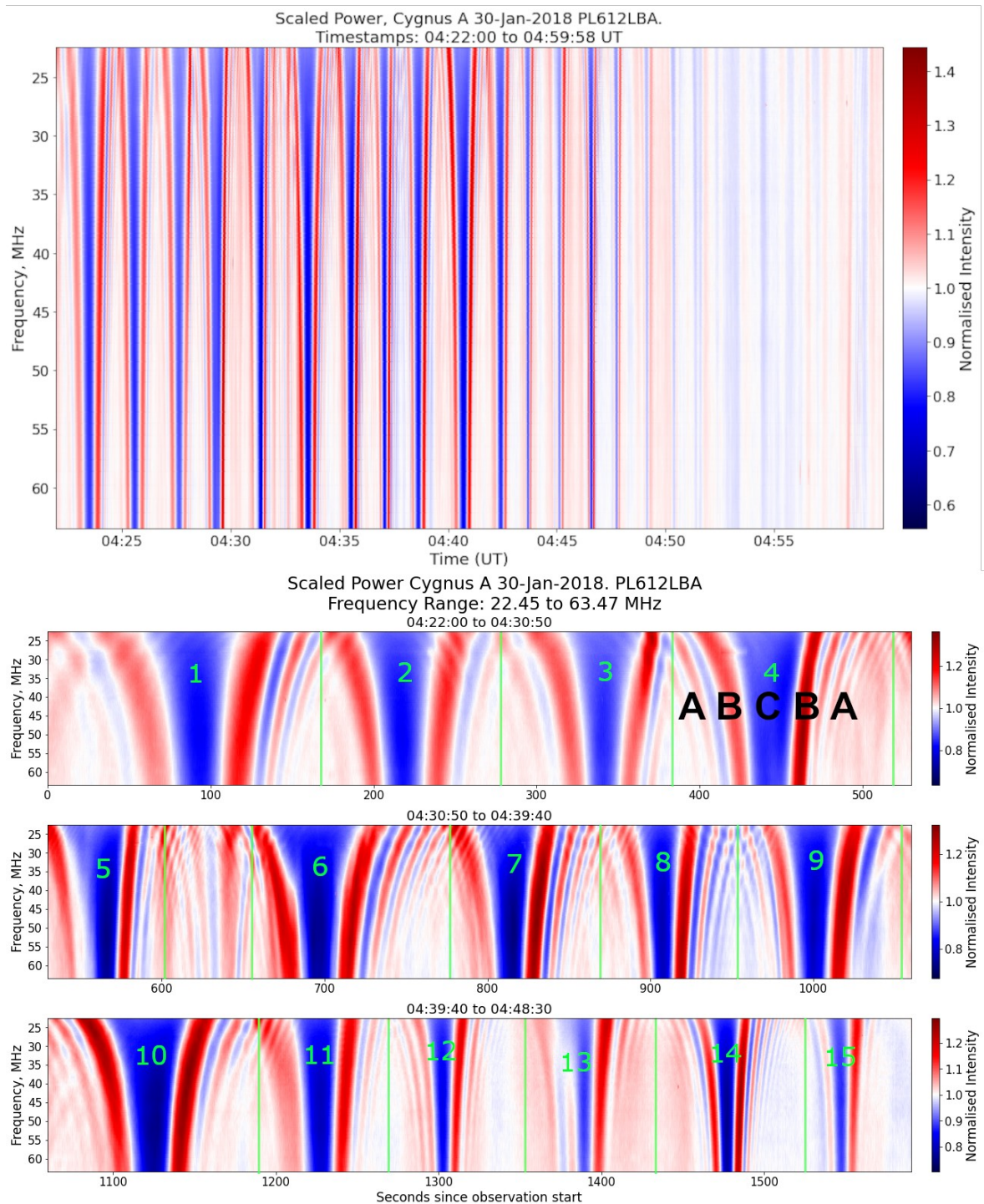

**Figure 1**: Top panel: Mean-centred dynamic spectra created from filtered data from the 40 minute observation of Cygnus A on 30[th] January 2018, showing lower frequencies at the top of the plot. Bottom panels: The dynamic spectra with the time axis scaled to better illustrate the fine structure in the features. Each individual feature has been numbered for further discussion. For reference, the regions common to each event labelled A B C we refer to in the text as 'secondary fringing / ringing irregularities', 'boundary enhancements' or 'boundary signals', and 'main' or 'v-shaped' signal fades, respectively. Signal intensity scales are the same in all plots.

Figure 1 clearly demonstrates the structure and strong frequency dependence generally present in each feature. Greater fine structure is more visible at lower observing frequencies. The major signal fades in each event are bounded on each side by clear ringing enhancement patterns, with a frequency-dependent curvature. There is also a subtle but repeated asymmetry seen when comparing the boundary signal intensities to the left and right of each numbered main fade, with the left boundary signal having a greater curvature and lower relative intensity. This is observed most prominently in features 12 and 13.

There are other differences between each of the features, and we can generally separate them into three groups. Features 1-4 are the most broad, resulting in less fine structure and fringing, particularly on the leading edge of the major signal fade. Features 5-11 show greater secondary fringing and are generally less broad, with a notable decrease in the relative intensity of the secondary fringing at higher frequencies (seen also in feature 14). Features 11-15 are characterised by distinctly narrower v-shaped fades, less definition in the fringes, and a greater asymmetry with regards to the intensity of the boundary signals on each side of the main signal fade. In some cases the secondary fringes of one event can be seen to overlap those of the next. This particular detail is only visible because these observations are broadband in nature. Analysis of this event with one or a small number of frequency channels, as is typical with many ionospheric scintillation studies relying on GNSS data (e.g. Kintner et al., 2005; Kintner et al., 2007; Song et al., 2022), would not have revealed this subtlety.

Figure 2 shows the dynamic spectra of the observation of Cassiopeia A made over the same time period at station PL612, again with numbered events. Cassiopeia A was at an azimuth of 24.5° and a moderately low elevation of 28.2° degrees at the start of this observation, rising to 27.6° and 29.8° for azimuth and elevation, respectively, by the end. Similar v-shaped fades and secondary fringing can be seen in the data as those in figure 1, however in this example we observe less well defined fringes, and stronger asymmetry in each of the ringing irregularity regions. Once again placing a lifetime on the event was not possible as it was already ongoing at the beginning of the observing window with fainter v-shaped fades still visible near the end.

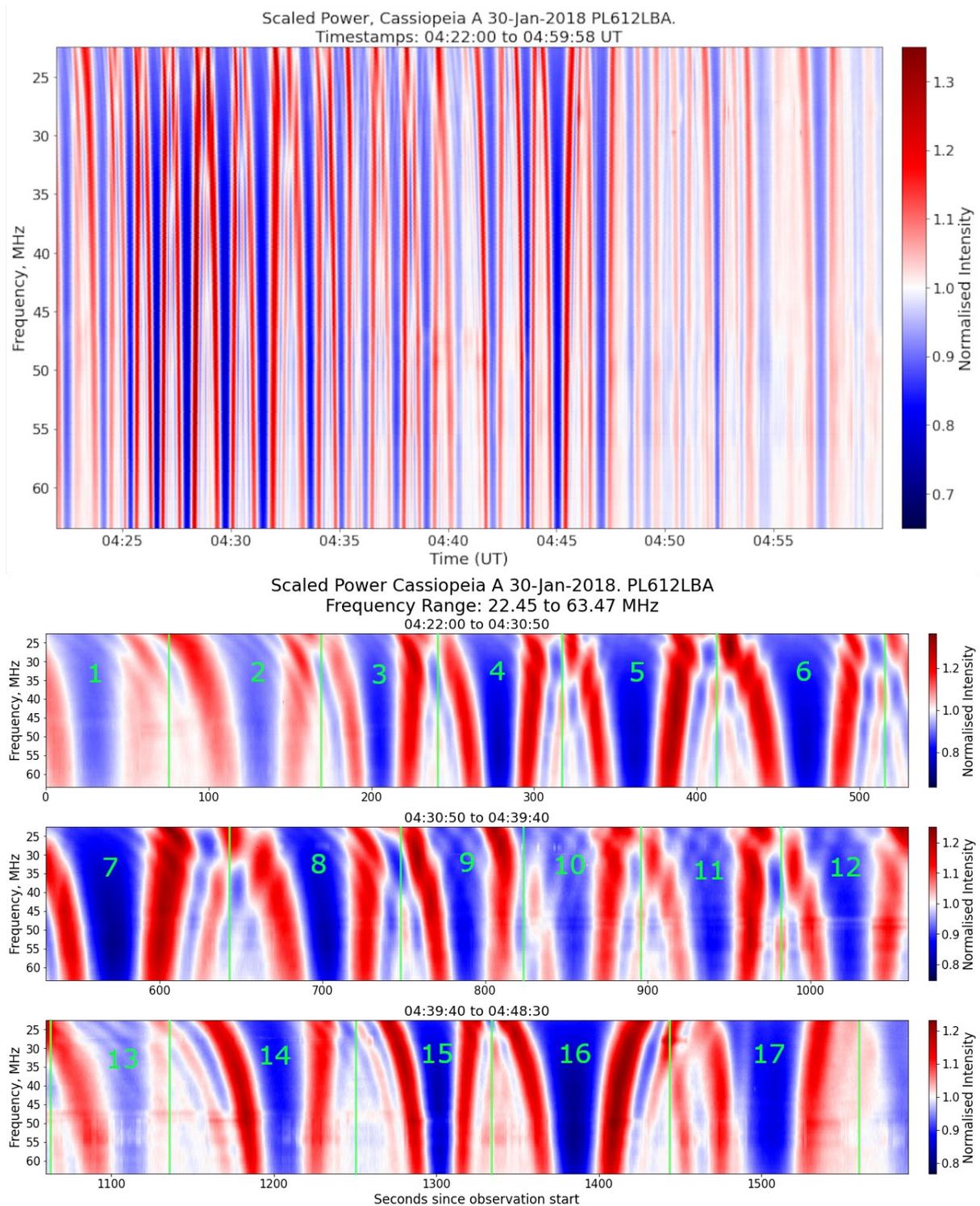

**Figure 2**: Top panel: Mean-centred dynamic spectra of normalised intensity data from the observation of Cassiopeia A by station PL612. Bottom panel: Zoomed in features, individually numbered. The bottom three panels are all plotted to the same time scale.

The asymmetry is particularly pronounced in events 1 and 13 in which, especially in event 13, the boundary signal enhancement to the right of the main fade is almost invisible. These events also overlap with each other more than they do in the Cygnus A observations, leading to the ringing enhancements from one event superimposed upon the main fade of the next one. This is especially visible in events 1-2 and 9-13, and at the lower end of the observing frequencies (~< 35 MHz). Estimating event onset time was not possible as the beginning of the observation period starts part way through the sequence of v-shaped fades.

**2.2 December 2016**

The second set of observations in which similar features appear is from 15 December 2016. As with the previous example, these observations were made using the two sources Cassiopeia A and Cygnus A; the dynamic spectrum comes from the LOFAR international station in the UK (UK608: latitude: 51.14°, longitude: -1.43°). This observing window ran from 18:18 UT to 19:49 UT.

The dynamic spectrum here has been time restricted to approximately the first 20 minutes of the full 1.5 hour observation period, as no features of interest were visible beyond that time. The bandwidth used here has also been restricted further to 22.5 – 60.9 MHz due to somewhat heavier RFI contamination at the previously used higher frequencies.

Unlike the 2018 observations using PL612, in this case the v-shaped signal fades with accompanying ringing irregularities were only detected in Cygnus-A observations and were not clearly identifiable in the Cassiopeia-A data, implying a fairly localised plasma scattering region. Data from several of the remote LOFAR stations in the Netherlands also showed evidence of the v-shaped fades however they were less well-defined than for the UK station; examples of these can be found in the supplementary material to this paper. At the start of the observing window the elevation and azimuth for Cygnus A as seen from UK608 was 49.4° and 278.8°, respectively. By the end of the observing window, the elevation of Cygnus A had decreased to 35.7°, and azimuth had increased to 293.0°

These observations show similar v-shaped signal fades again bounded on each side by boundary signals and ringing irregularity regions, but with no significant overlap between events and much greater detail of the individual fringes in the ringing irregularity regions being visible. Figure 3 shows the dynamic spectra from UK608 observations of Cygnus-A. We observe similar asymmetric variations as seen in the 2018 data, with the right side fringing of the numbered events significantly more intense. However, the overall form of each event remains highly symmetrical. The ringing irregularity regions, particularly for event 3, are very well defined with >10 individual secondary fringes visible after the main signal fade, and especially at the lower observation frequencies.

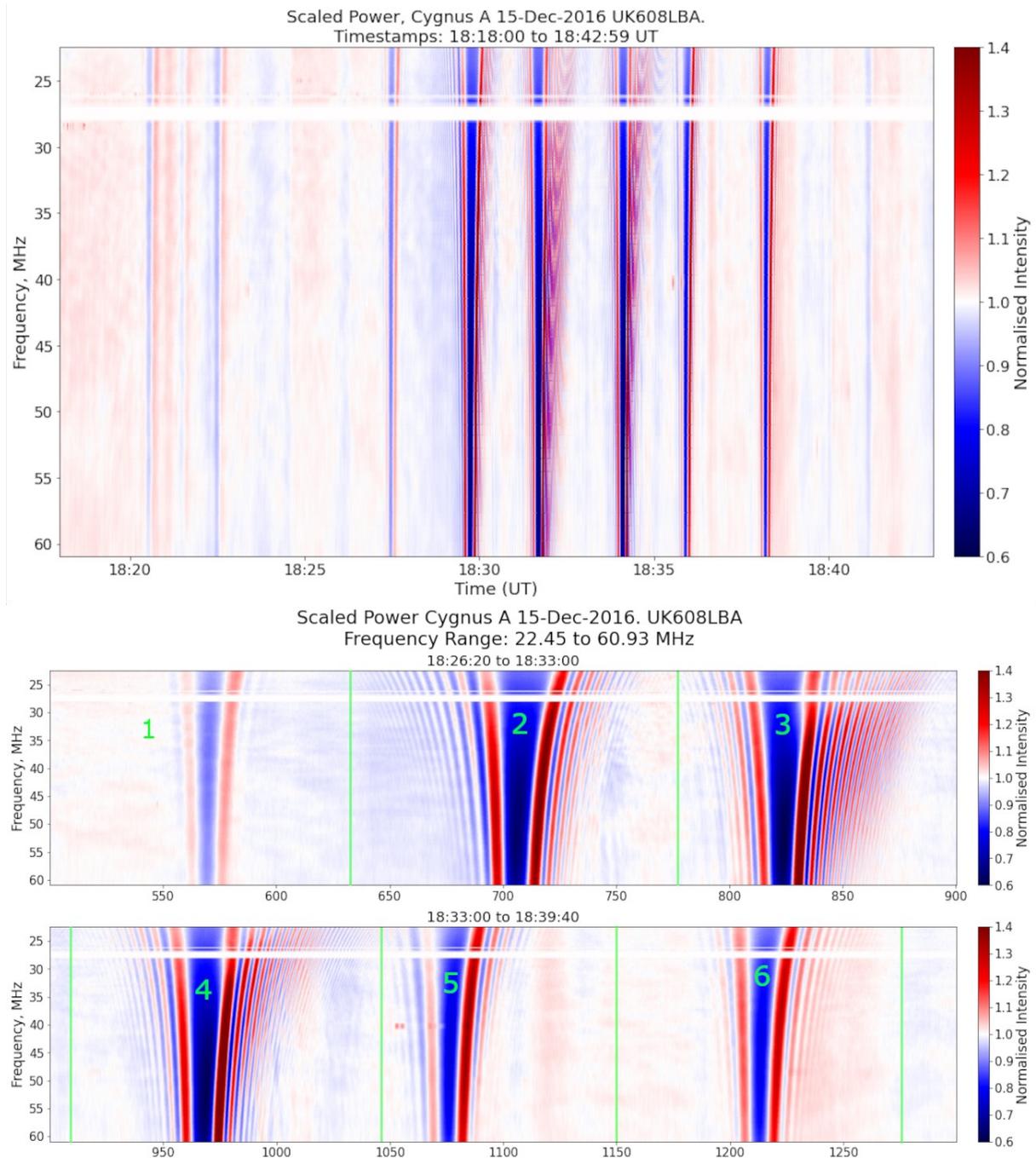

**Figure 3**: Top panel: Mean-centred dynamic spectrum of the observation of Cygnus A by LOFAR station UK608. Bottom panel: Colour scale corresponding to normalised intensity has been capped at a reduced value of 1.4 so that the lower intensity fringing in events 1 and 6 are more visible. All plots use the sam intensity scale, and the bottom two panels are plotted to the same time scale. The horizontal white streaks are channels where RFI which has been removed.

## 3. Data analysis

We analyse these observations in several stages. Firstly, in section 3.1, the observation geometry is examined. In section 3.2, we utilise delay-Doppler spectral analysis to estimate the propagation velocity of each event. Finally in section 3.3 we examine the periodicities of each case study and attempt to estimate the separation distances between each individual plasma structure based on the estimates for propagation velocity.

### 3.1 Observing Geometry

To attempt to establish a propagation altitude for each event we examined data from ionosondes located close to ionospheric pierce points (IPP) for the LOFAR stations in each case study. Ionosonde data from FF051 (Fairford, UK, 51.7°N, 1.5°W) and RL052 (Chilton, UK, 51.5°N, 0.6°W) were available for the 2016 observations, but unfortunately suitable data was not available from ionosondes close to the IPP for the 2018 observations. Figure 4 shows four ionograms taken on 15[th] December 2016, during the 2016 observations. Two are from FF051 at 18:15 and 18:30 UT in the left column, and two are from RL052 at 18:00 and 18:40 in the right column. The ionograms from FF051 have 15-minute resolution and the ionograms from RL052 have 10-minute resolution.

The two top panels show evidence of semi-blanketing sporadic-E at an altitude of ~110 km, in which some portion of the radiated energy from the ionosonde passes through the E-layer with sufficient intensity that echoes from the F-layer can also be seen. The F-region traces are still visible in the same frequency range but the backscattered energy appears partially absorbed. The bottom two panels show, aside from what is likely noise at the lowest altitudes, only F-region with the altitude peak of this layer, hmF2, at ~295 km in both. Sporadic-E seen in the earlier ionograms apparently has dispersed by this time. The F-region traces, now fully visible as a result of the removal of the semi-blanketing E-layer, show little evidence for spread-F or trace bifurcation which would indicate the presence of a travelling ionospheric disturbance in the field-of-view (Bowman et al., 1987; Moskaleva & Zaalov, 2013; Jiang et al., 2016). Furthermore, when the sporadic-E is present, particularly in RL052 (figure 4, top right panel), it displays altitude spread, indicating that it is not simply a uniform thin ionisation layer.

From & Whitehead (1986) investigated several types of E-layer structures and demonstrated that spread-E and semi-blanketing or partially-reflecting sporadic-E is a structured medium consisting of clouds of electrons. By contrast, fully-blanketing or fully reflective sporadic-E is caused by thin sheets of more uniform ionization. The LOFAR observations clearly imply structured plasma. Furthermore, the periodicities (see section 3.3) of the observed features varies between 75-125 seconds, which is much closer to the Brunt-Väisälä frequencies for altitudes of 100-120 km (<5-minutes) than for F-region altitudes (>10-minutes; e.g. Snively & Pasco, 2003; Borchevkina et al., 2021). Given the absence of strong evidence for perturbations to the F-region in the ionograms, the events seen in the 2016 data are most likely propagating in the E-region, which, unlike the F-region ionogram traces, clearly shows evidence for structured plasma at two different ionosondes at multiple times throughout the observing window.

We note that the positions of the IPPs at event time in the 2016 data do not exactly overlie the ionosondes. Therefore one should not expect an exact time match between structures seen in

LOFAR and structures seen in the ionosondes. The picture of the ionosphere they give, in terms of the LOFAR data, is an approximation of overall conditions in the region.

The absence of any such contemporary data from the 2018 observations however means we cannot definitively argue that the 2018 event was located in the E- or F-regions. Instead, we proceed with the analysis for that case on the basis that both possibilities may be true and use the best available estimates for altitude.

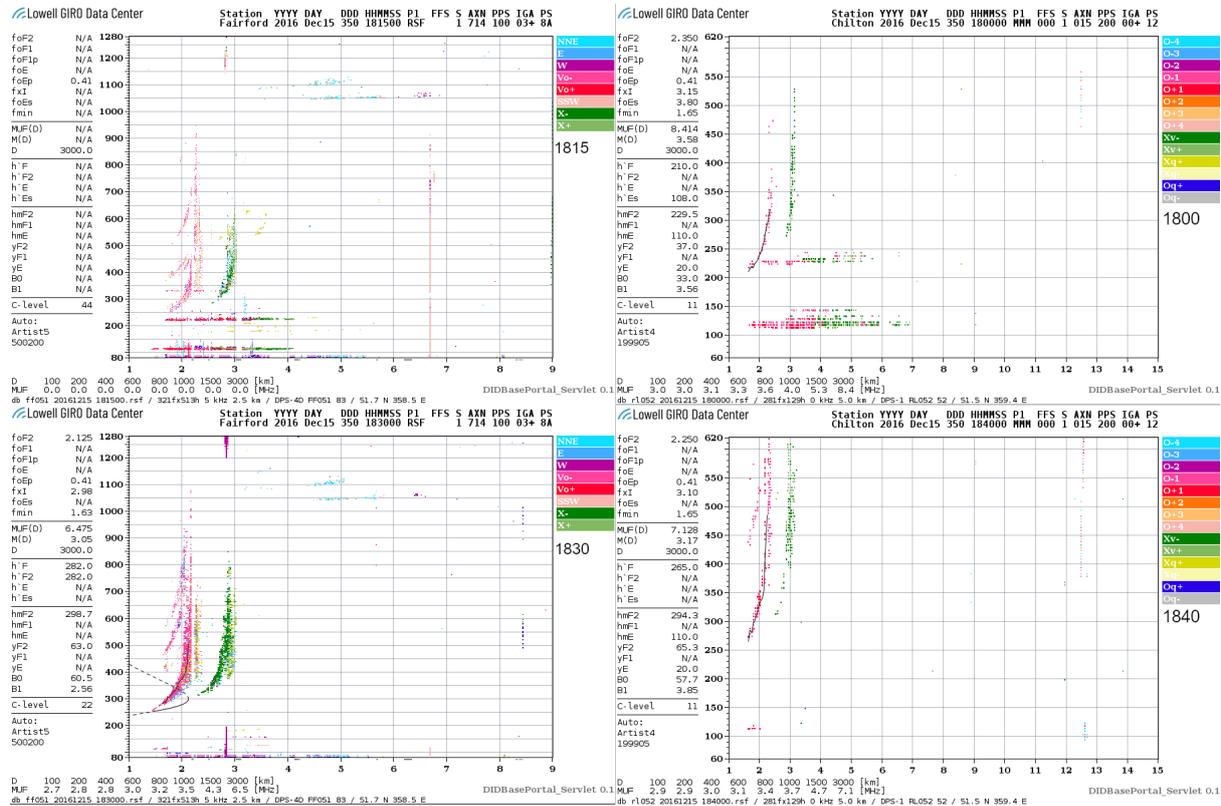

**Figure 4**: A set of ionograms from the Fairford (left column) and Chilton (right column) ionosondes taken on 15th December 2016 at the times shown, to coincide with the LOFAR observations shown in figure 3. Little evidence of any disturbance is visible in the F-region, however non-blanketing sporadic-E is observed intermittently in both ionosondes. The bottom right panel in particular has a maximum useable frequency of < 3 MHz, indicating a particularly low plasma density in the F-region.

An estimate was made of the approximate height of both layers using measurements of the average hmE and hmF2 from the Fairford and nearby Chilton ionosondes (located at latitude: 51.70°, longitude: 358.50° and latitude: 51.50°, longitude: 359.40° respectively). This gives the E layer peak at 110 km, and the F layer peak at 297 km.

In figure 5 IPP maps are shown for the 2016 observations using the UK and remote LOFAR stations. In this case events were only detected in Cygnus A data; IPP arcs for Cygnus A are shown in cyan. Even though it seems more likely that the E-region is the appropriate propagation altitude IPP arcs for 110 km and 295 km are presented for comparison. The position of the Fairford ionosonde is also shown. Furthermore, the approximate position and timing of each event as seen on each station is indicated by the yellow spots and accompanying UT timestamps. The IPP projections are calculated using the spherical Earth approximation method described in Section 3 of Dorrian et al., (2023), based on the geometry outlined in Birch et al., (2002). A full set of v-shaped fades consisting of a clear onset and end time set against an undisturbed background ionosphere was seen only from UK608; with partial detections either beginning or ending part way through a series of fades at several of the remote stations. Combined with the fairly rapid appearance and disappearance of non-blanketing sporadic-E in the ionograms, this suggests that, rather than some singular feature like a travelling ionospheric disturbance wave passing through the area, there was instead a regional filamentary sporadic-E structure within which multiple QPO generating structures were propagating with similar characteristics.

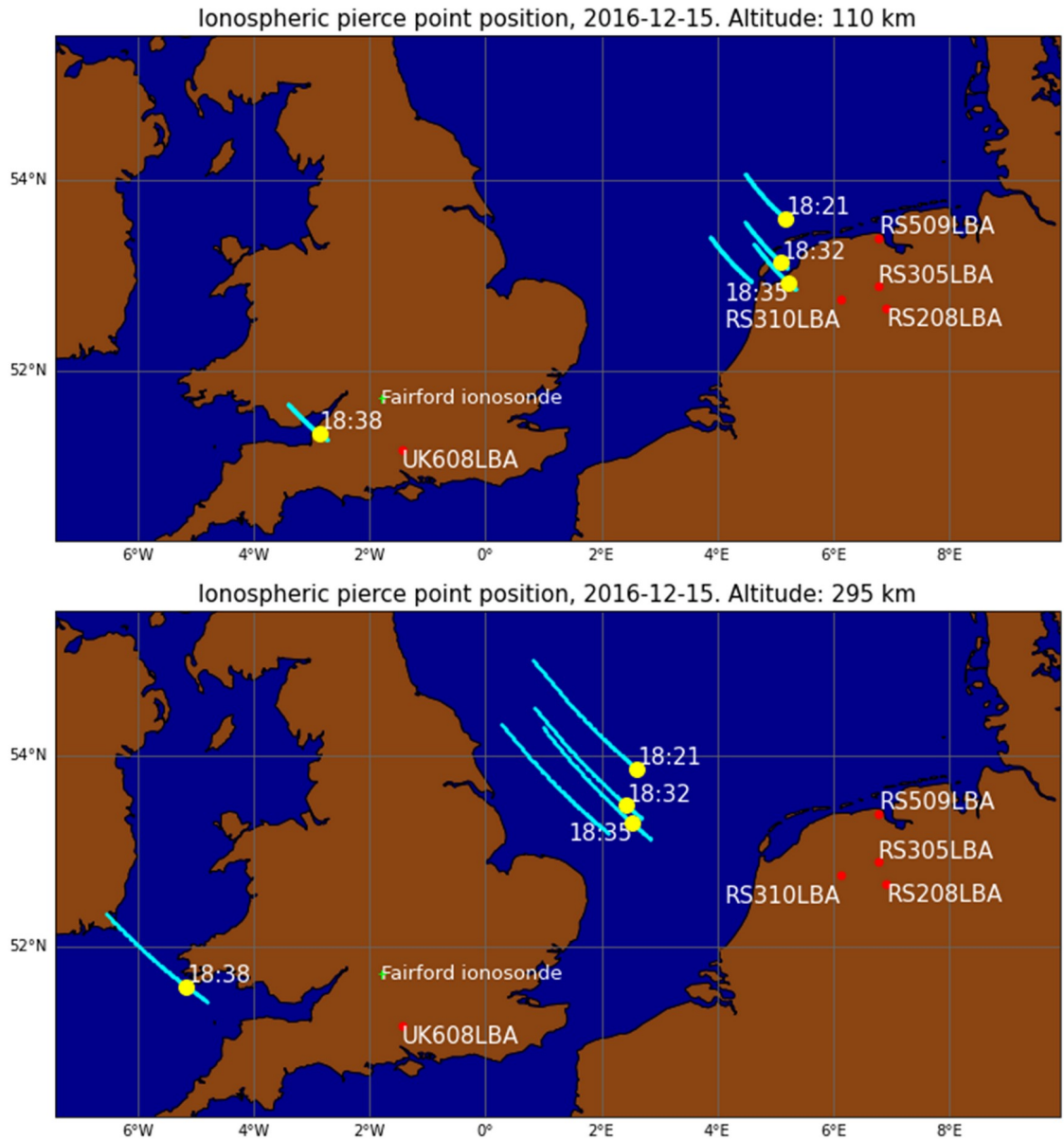

**Figure 6:** Map of IPP location for the 2016 observation of Cygnus A. Shown on the map is the UK station and several of the remote stations (prefix RS) in the Netherlands in which the event appears.

The absence of useable ionosonde data from the 2018 event restricts the analysis to considering that the propagation was either E- or F-region. It is also highly unlikely to be a

D-region phenomena given that the observations were made pre-sunrise in winter when the D-region is typically less prominent (e.g. Renkwitz et al., 2023). As previously mentioned, the geophysical conditions during this observation were very quiet. Consequently we use estimated E- and F-region altitudes from the 2016 IRI model (Bilitza et al., 2016) for this locale which yielded hmE and hmF2 altitudes of 110 km and 250 km respectively. In Figure 6 the left and right panels show the calculated positions of the IPPs for altitudes of 110 km and 250 km, respectively, for both Cassiopeia A (orange arc) and Cygnus A (cyan arc). The start and end times of the observing window are shown as is the position of LOFAR station PL612 which recorded the dynamic spectra shown in Figures 1-3. That the QPO essentially fills the observing window on both sources prevents us from assigning any maximum constraints as to its total lifetime, other than to state that it lasted for a minimum of 37 minutes (i.e. the length of the observing window). If one assumes, as seems likely, that the events seen on both radio sources are generated by the same overall ionospheric regional structure then, assuming an E-region altitude of 110 km, this regional structure would need to have a horizontal size of at least 115 km to be present at both IPP arcs simultaneously. Likewise, if one assumes an F-region altitude of 250 km, the horizontal size would have to be a minimum of 300 km.

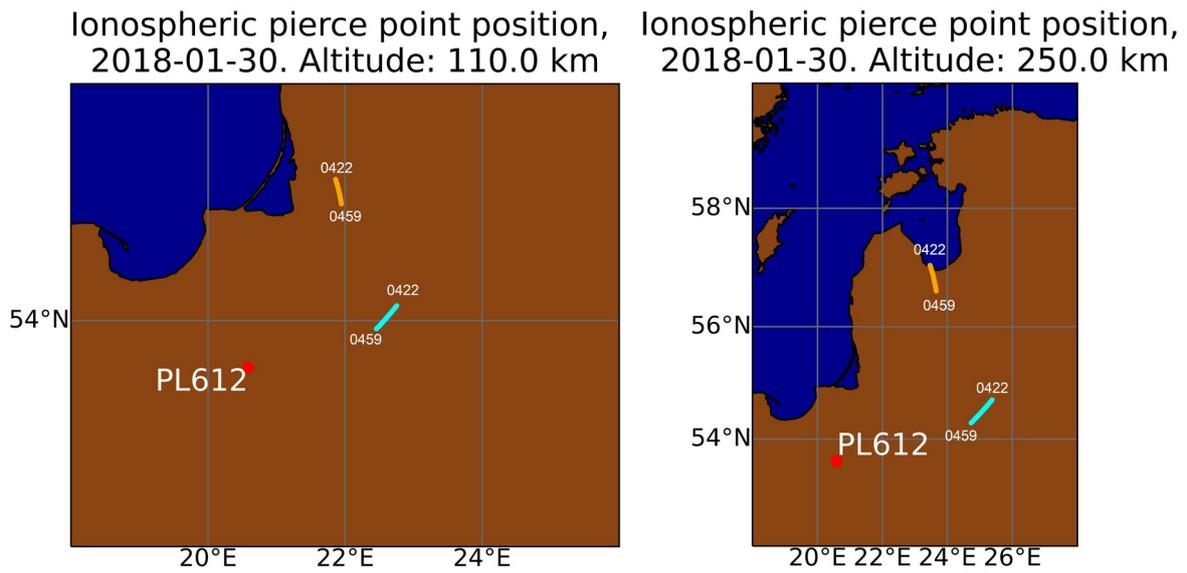

**Figure 5**: Maps of the position of the ionospheric pierce point (IPP) for the 2018 observations of Cygnus A (orange arcs) and Cassiopeia A (cyan arcs). Shown on the left is the estimated location of the IPP for the E-layer at 110 km, and on the right for the F layer at 250 km.

### 3.2 Delay-Doppler spectra and velocity estimation

The next stage of analysis was the creation of the delay-Doppler spectrum (DDS), from the primary data in the dynamic spectra. The DDS's are the 2-D fast Fourier transforms of windowed sections of the dynamic spectra, and their primary utility here is the determination of propagation velocity. A DDS can exhibit characteristic scintillation arcs, the curvature of which is a function both of the propagation velocity of the plasma structures and their distance from the ground station antennas. Hence, if one can isolate either of these characteristics then it is possible to extract the other (Cordes et al., 2006) using equation 1,

$$L = 2CV^2 \qquad \text{Eq. 1}$$

where L is distance along the line-of-sight to the scattering region, V is the plane-of-sky velocity component of the moving scattering screen, and C is the curvature of the scintillation arc.

Extracting L or V, with the intention of isolating the other, can be accomplished with the use of contemporary data from, for example, ionosondes. Such arcs have been used in the field of interstellar scintillation for some time (e.g. Stinebring et al., 2001) but, in an ionospheric context, using artificial satellite radio sources first reported by Cannon et al., (2006), and first used on natural radio sources, again in an ionospheric context, with a multi-octave bandwidth, by Fallows et al., (2014).

The arcs arise partially as a consequence of the Huygens principle in which a plane wave, which is considered as a summation of individual wavelets, is incident upon a scattering screen, in this case ionospheric plasma. This generates an ensemble of secondary spherical wavelets, which undergo mutual interference as they propagate forwards from the screen. Spherical wavelets propagating from the screen at larger distances from the observer will exhibit a greater signal delay time than those which propagate from positions on the screen nearer to the observer, as a result of the longer path lengths they must take. If there is a non-zero relative velocity between the scattering screen and the observer position then the same wavelet ensemble undergoes Doppler shifting, as a function of the observers viewing angle to the scattering screen. The frequency shift is minimised when observing the point on the scattering screen closest to the observer, and increases as the viewing angle increases. The resulting arcs are thus a convolution of both the variable path length taken by each wavelet and the Doppler shift resulting from the relative velocity of the screen with respect to the observer. Multiple arcs appearing in the same DDS are also of interest as different individual arcs imply that there are different populations of scattering plasma in the raypath, and offer a means of separating the velocities of each population. A detailed synopsis of scintillation arc formation can be found in Cordes et al., (2006).

*2018 data*

To isolate the curves from each of the periodic features, the data were sliced into the numbered regions defined in section 1. Each DDS in figure 7 was created using the dynamic spectrum from the PL612 observation of Cygnus A in 2018. The vertical axis in each DDS is the quantity β (in units of $m^{-1}$) which is the conjugate to observing wavelength. The weaker definition of the ringing irregularity regions in the Cassiopeia A data from this observation precluded the formation of clear arcs. Each sub-plot shows the dynamic spectrum for a single feature in the numbered sequence from figure 1 and the corresponding DDS. Clear scintillation arcs can be observed in all cases as can a variation in definition and extent in β.

Parabolas can be fitted to the most well-defined arcs in order to find their curvature, however fitting the curves automatically is technically challenging as any code must understand which data points are part of the arc and which are not, with ambiguities having the potential to significantly affect the curvature estimate (Fallows et al., 2020). Due to the large number of curve fits performed, it was necessary however to utilise an automated routine. Points in the secondary spectrum with a power above a threshold of approximately 60 dB were sampled to isolate the parabolic arc. Arc identification for the code was then simplified by restricting the region of the DDS used for the fit, excluding those regions near the horizontal and vertical axes which are often contaminated by noise. The remaining data points were fitted to a parabola of the form $y = Cx^2 + Bx$ using the least squares method to optimise C and B. While the term in B was used in the fit to account for a shift in the image of the radio signal due to larger scale phase gradients (Cordes et al., 2006), it was not included in the calculation of velocity from curvature.

Moreover, several of the DDS exhibit secondary scintillation arcs which lie closer to the vertical axis than the primary arc. Furthermore, secondary arcs often have a different curvature from the primary arcs. Where clearly identifiable secondary arcs are found, arc curvature has been fitted to these also and, for clarity, velocities extracted in the analysis of these secondary arcs is identified as such. Further complications occasionally arise when the DDS scintillation arcs are asymmetrical, sometimes with the arc being visible only on one side of the central frequency.

The arc fitting process was performed on as many DDS's as possible given the limitations on arc definition described above. Furthermore, for each event, two velocity values are extracted, one assuming an E-region propagation altitude and the other assuming an F-region propagation altitude. The results of this process are shown in figure 8, with red and blue crosses indicating whether the curve fit was performed on a primary arc, or secondary scintillation arc where one was observed. Note, only Cygnus A observation data was used for this process, as the weaker secondary fringing definition in the Cassiopeia A data suppressed any consistent scintillation arc formation.

A general increase in velocity of approximately 70 ms$^{-1}$, from 50 ms$^{-1}$ to 120 ms$^{-1}$, is observed when using the E-region altitude scaling. If one uses the F-region scaling then the velocity increase is ~120 ms$^{-1}$, rising from 80 ms$^{-1}$ to 200 ms$^{-1}$. From equation 1 it can be seen that if a constant propagation velocity were assumed, then changes in curvature could only be explained by rapid changes in altitude between one v-shaped fade and the next in the sequence. Instances of primary and secondary curves being seen in the same DDS are likely due to more than one population of scattering plasma in the raypath, with each propagating at its own velocity, and both contributing to the overall signal. Once again it is noted that only because these observations are broadband, that such a characteristic may be extracted.

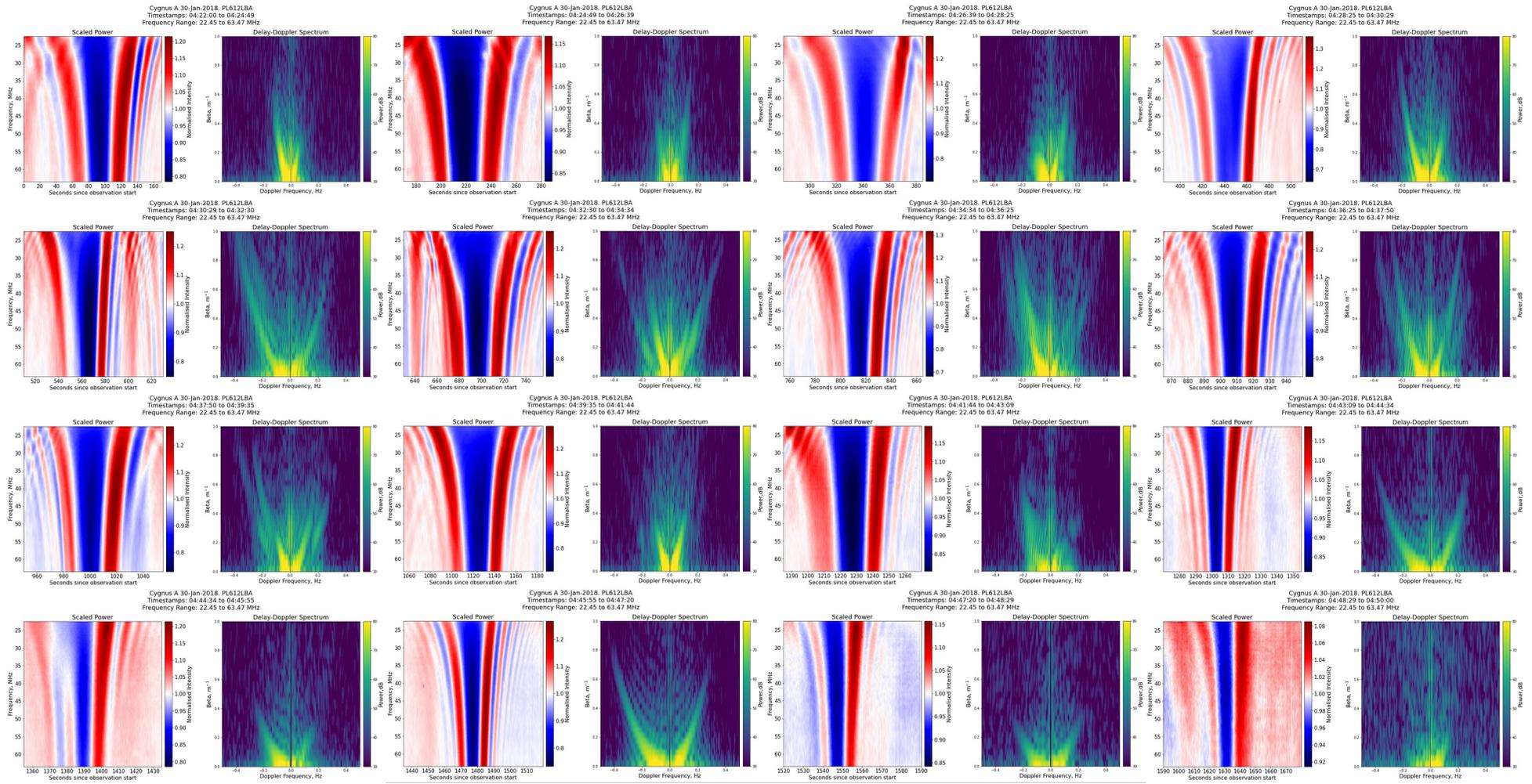

**Figure 7**: The set of numbered features from figure 1 (30 January 2018 Cygnus A dynamic spectra from PL612), alongside their corresponding DDS. Scintillation arcs are clearly visible in all cases and vary considerably during the observation.

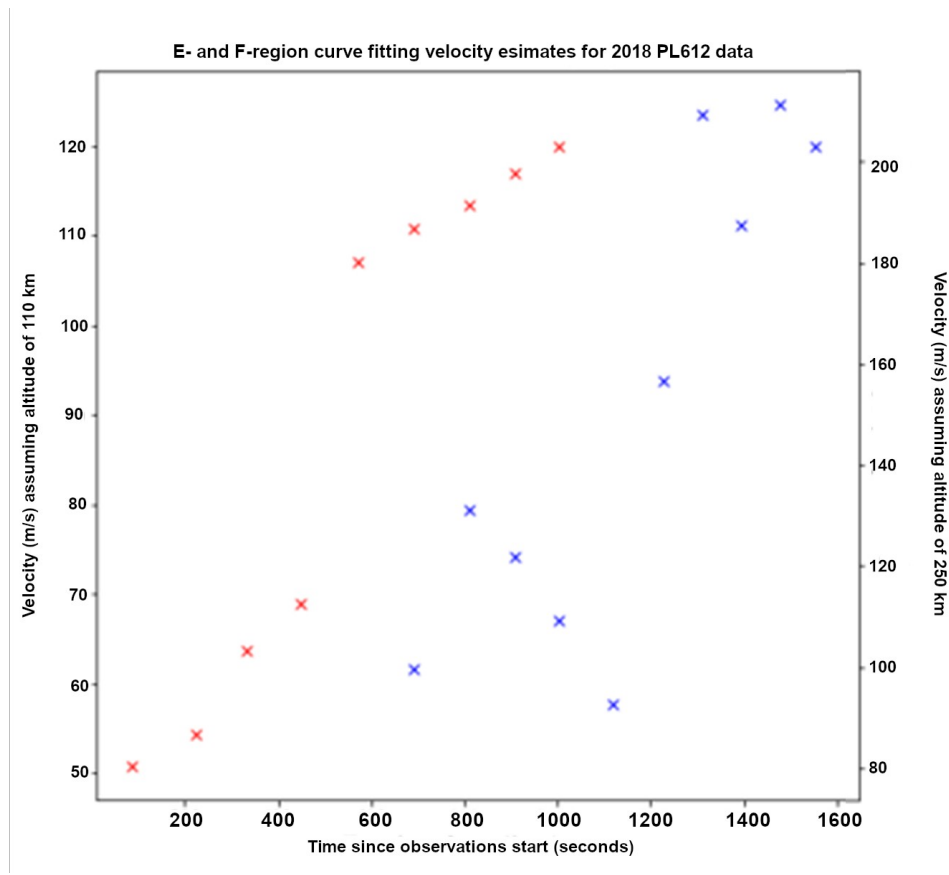

**Figure 8.** Red crosses denote velocity estimates from primary curve fits, blue crosses denote velocity estimates from secondary curve fits where applicable. The left hand vertical axis shows the velocities if E-region propagation altitude is assumed (110 km), and the right hand vertical axis shows the velocities if F-region propagation altitude is assumed (250 km).

*2016 data*

The same process was then performed on the 2016 data. In figure 9 the DDS's constructed using the 2016 data from UK608 are presented in the same format as figure 7, with each numbered feature being associated with its corresponding DDS. A combination of the events in the 2016 data being separated in time such that event overlap is minimised and, in several cases, clearer definition of secondary fringing, gives the scintillation arcs in these cases much clearer definition. The sensitivity of arc formation to the presence of well defined secondary fringing is clearly observed in the lower arc definition in the first panel (top left) in figure 9 in which the dynamic spectra for that particular event exhibit weak or no fringing, whereas in all the others it is clear. The consistent asymmetry in the arcs throughout is a reflection of the larger number of clearly defined fringes to the right of the main signal fade as well as intensity differences between boundary signals.

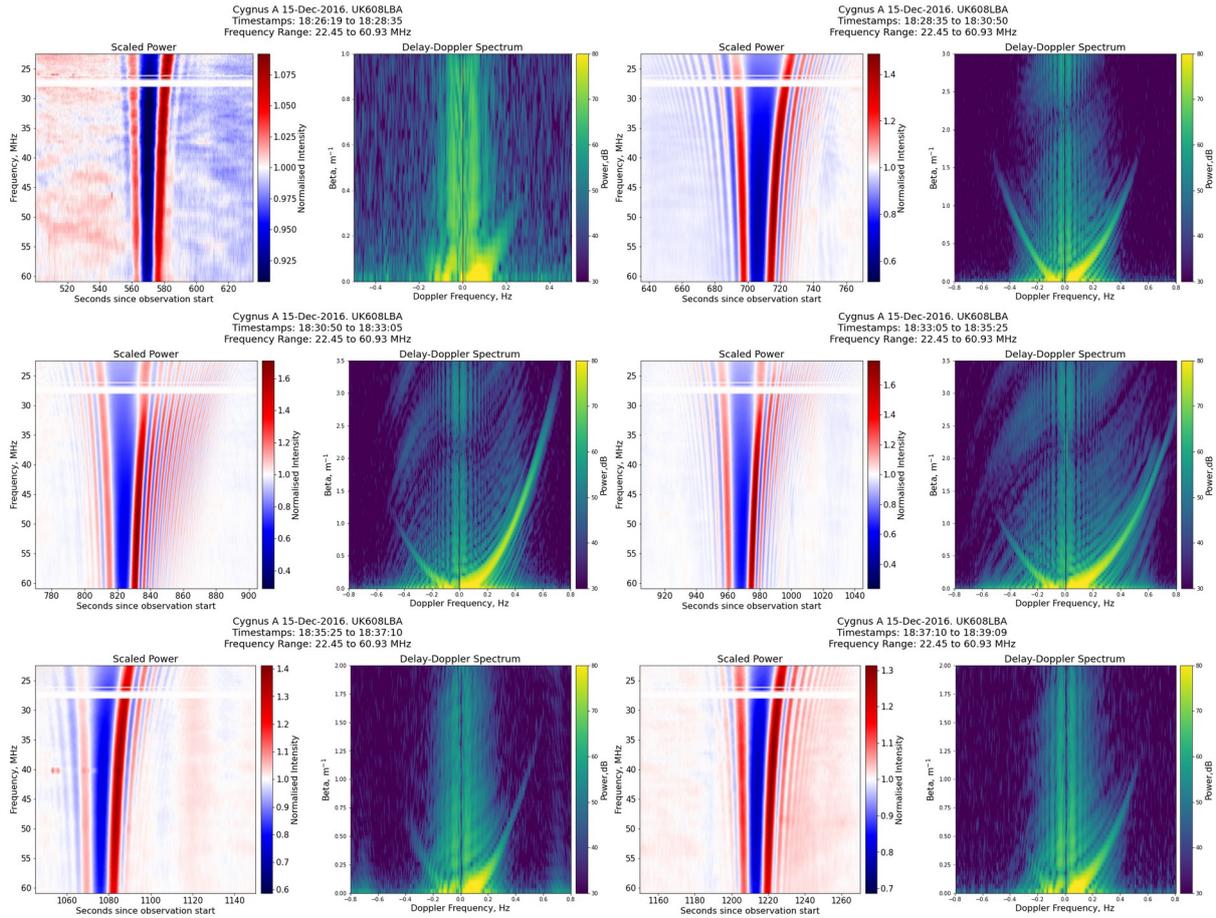

**Figure 9.** Individual events from the dynamic spectra recorded on UK608, Cygnus A observations on 15 December 2016, and their corresponding DDS. Asymmetry in the arcs is also a consequence of secondary fringing, but in this case, being due to the clearer definition of fringing.

Figure 10 shows the results of the velocity analysis using the 2016 data, from all stations in which the QPO is seen. These include the UK station (UK608) and several of the remote stations in the Netherlands (RS208, RS305, RS408), with the different colours indicating different ground stations as shown. Curve fitting for these data was simpler than for the 2018 data as the definition of DDS scintillation arcs was much improved. Additionally, F-region propagation is discounted given the absence of spread-F or trace bifurcation in the ionograms which would otherwise support perturbations in the F-region. Unlike the analyses of the 2018 data, in this case a consistent propagation velocity range of between 110-130 ms$^{-1}$ is seen throughout, with no indication of any plasma screen acceleration.

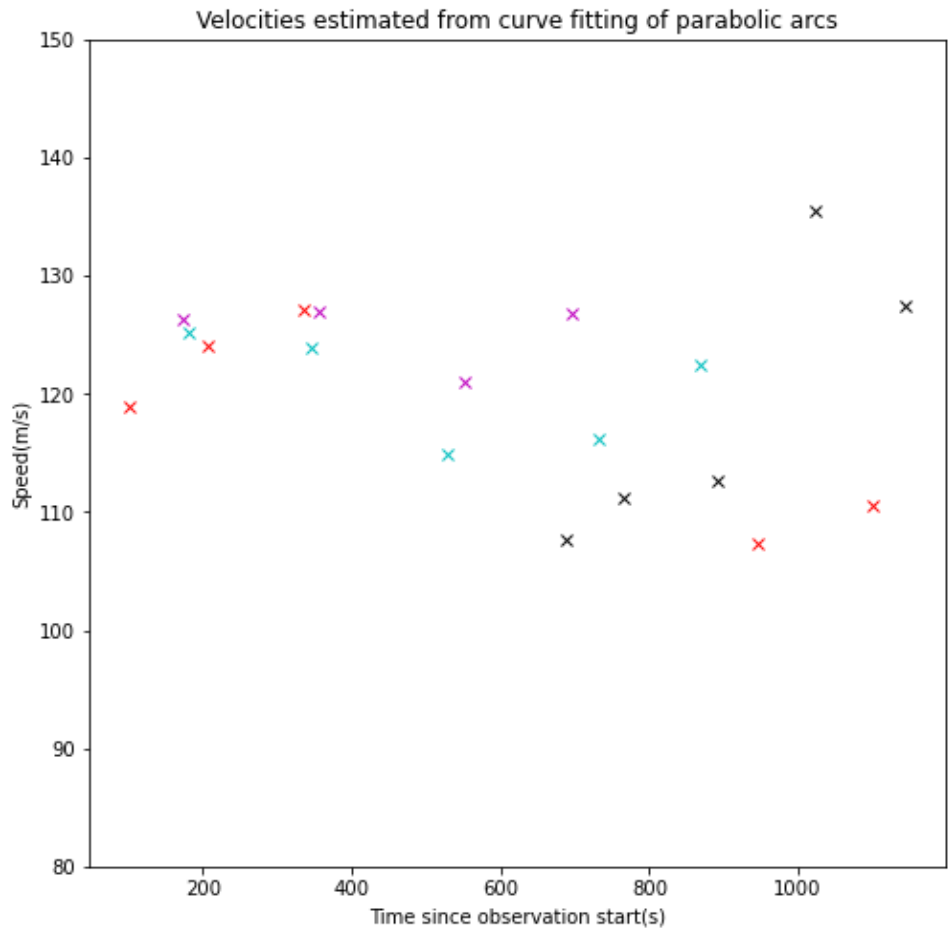

**Figure 10.** Velocity estimates from curve fitting for the 2016 observations of Cygnus A using an E-region propagation altitude of 110 km, with different point colours indicating different LOFAR stations. Black crosses are for UK608, red for RS208, cyan for RS305, and magenta for RS407. No F-region scaling has been performed in this case given the stronger evidence for event propagation in the E-region.

### 3.3 Periodograms

With some estimates of propagation velocity now established we can proceed to estimate the separation between sequential v-shaped fades in the scattering screen by investigating their periodicities. To do this, periodograms were constructed from each frequency channel using Welch's method (Welch, 1967). The periodicities from the 2018 observations using data from PL612 on Cygnus A are shown in figure 11 (top row) and Cassiopeia A (bottom row). The periodograms are taken from time-windowed sections of the dynamic spectra as indicated by the time stamps shown. This was due firstly to the fact that, in the 2018 observations, a distinct increase in velocity was seen over the course of the observing window. As velocity is key to extracting distances between each plasma structure, it was therefore important to separate the data into the given sections to avoid smearing out the periodicities. Secondly, the time lengths of the individual v-shaped fades changes throughout the observing window.

The time sections analysed run from 04:22-04:30 UT corresponding to events 1-4 in Figure 1, from 04:30-04:41 UT corresponding to events 5-11, and finally 04:41-04:48 UT corresponding to events 12-15. The breaks in each window coincide with broad changes in characteristics of the v-shaped signal fades with 1-4 being quite broad and with somewhat weaker secondary fringing definition. The v-shaped fades 5-11 are consistently of a similar time length and show collectively more secondary fringing. Finally, there is a notable change in width (and hence lifetime) of the v-shaped fades between events 11 and 12, with events 12 and beyond all having distinctly narrower v-shaped fades. Velocity estimations from 2018 come only from the Cygnus A observations, however we also present the periodicities from the Cassiopeia A data for comparison.

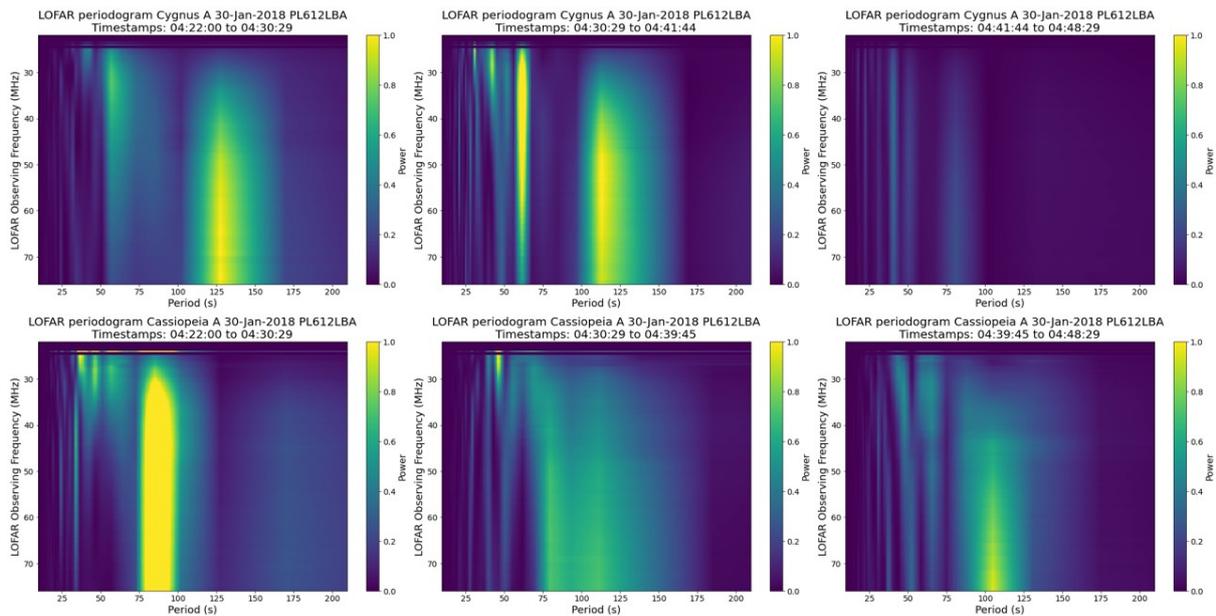

**Figure 11**: 2D periodograms for Cygnus A (top row) and Cassiopeia A (bottom row) observations on 30th January 2018 using LOFAR station PL612, selected for the time windows as shown. Intensity and time scales are the same for all plots.

Clear peaks in periodicities are observed between 75-125 seconds with several periodicities at shorter time scales. The main broader peaks are the periodicities of the v-shaped signal fades, whilst the shorter time peaks are caused by the secondary fringing. Peak strength varies throughout the observation window considerably. As shown in figure 9, the propagation velocity increases substantially during the passage of the event through the raypath, accelerating from ~50 ms$^{-1}$ to 120 ms$^{-1}$, assuming an E-layer altitude, and from ~80 ms$^{-1}$ to 200 ms$^{-1}$, assuming an F-layer altitude. If these velocities are converted using an average periodicity of 100 seconds into scale sizes then we are observing spacing between sequential plasma structures of 5-12 km in size assuming an E-region altitude, and 8-20 km if we assume an F-region propagation altitude.

Figure 12 shows the periodicities recorded from the 2016 Cygnus A data on UK608. Here, given the much shorter event life time and only 6 discernible individual structures, and the more consistent velocity estimation across multiple stations, the periodogram has been calculated over the entire dataset without time-windowing. Again a similar pattern is observed, with a main peak in periodicity at ~120 seconds, with notable secondary peaks at shorter timescales of 20-50 seconds. Again, this is due to the main peak being dominated by the main v-shaped signal fades, whilst the shorter periods are from the secondary fringing. Note also, both in figures 11 and 12, how the frequency dependent behaviour of the secondary fringes is reflected in the periodograms, with much clearer periodicity peaks at the lower end of the observing frequency range. Again, this is a characteristic which could only be extracted using broadband observations.

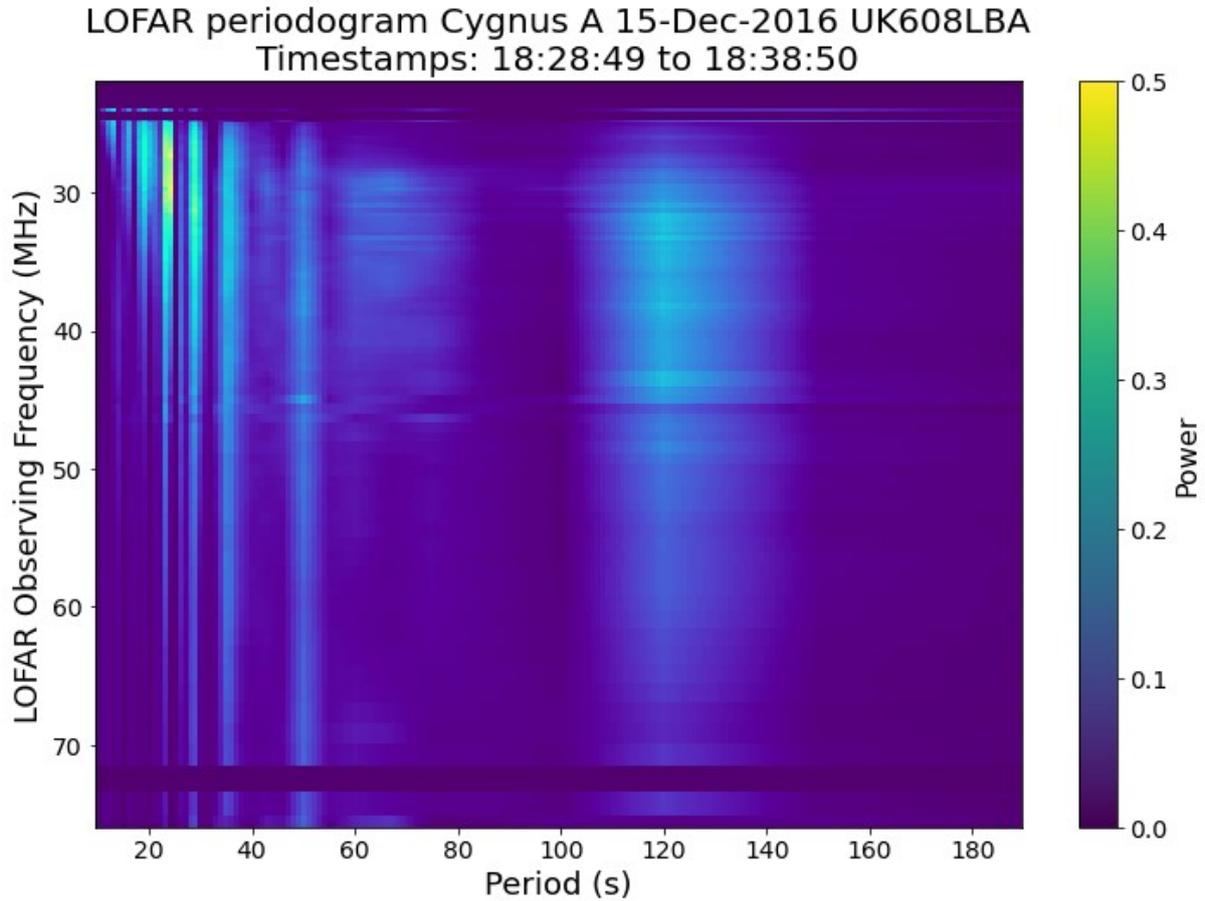

**Figure 12.** 2D periodogram for observations of Cygnus A in UK608, 15th December 2016. The main v-shaped fade periodicity peaks at 120 seconds, with shorter periodicities for the secondary fringing seen at <60s.

Estimates for velocity from the curve fitting for the 2016 data consistently yielded values of ~ 120 ms$^{-1}$, over several LOFAR stations. If one converts the periodicity peak of ~120 seconds, using this velocity, then in this case the individual plasma structures causing the v-shaped fades are separated in distance by ~14 km. These are approximately the same dimensions as the distance between the primary v-shaped fades in the 2018 case.

# 4. Modelling

To further assess the characteristics of the plasma structures we have attempted to reproduce them using a Gaussian thin screen phase model of the ionosphere. This approach was recently applied very successfully to spectral caustic lensing features in the ionosphere by Boyde et al. (2022). The model can be tuned with a variety of plasma scattering characteristics such as wave propagation velocity, background plasma density gradient, scattering source size and shape, and altitude (or distance from observer). The main output of a given model run is a synthetic dynamic spectrum which can be compared to the original observation.

Treating the ionosphere as a thin phase screen (or several screens) is widely used for modelling ionospheric radio propagation (e.g. Carrano et al., 2020; Hocke & Igarashi, 2003; Meyer-Vernet, 1980). For certain phase screen perturbations, it is possible to derive an analytic solution for the observed intensity distribution as a function of time (e.g. Meyer-Vernet, 1980). However, in most cases a numerical solution is required, the mathematical framework for which is described by Sokolovskiy (2001). The phase screen approach was first applied to replicating LOFAR data by Boyde et al. (2022), building on the earlier theoretical work of Meyer-Vernet (1980). As the solution is derived as a spatial rather than temporal intensity distribution, a constant velocity must be assumed to obtain a dynamic spectrum to compare to observations. The amplitude of the phase perturbation applied by the screen is assumed to be inversely proportional to the radio frequency, which is a valid approximation provided the local plasma frequency remains well below the radio frequency. The ionograms in figure 4 suggest a peak plasma frequency of ~3-4 MHz, suggesting that any deviation from this approximation will be negligible except possibly at the lowest observing frequencies.

In the model runs presented here, the source brightness distribution is assumed to be Gaussian and the progression and spacing of the Gaussian ionospheric perturbations as they move through the simulated raypath have been fixed based on the timestamps used to separate the features in section 2. The amplitude of each Gaussian (in terms of phase change), and the standard deviation have been adjusted by eye to match the intensity variation, asymmetry, and fringing present in each of the features. The perturbation amplitudes in the model are expressed in rad Hz, and a perturbation of $10^{10}$ rad Hz corresponds to roughly 1 TECu of line of sight TEC perturbation. The broadening of the v-shaped signal fades, and the density of the secondary fringing is partially a function of the velocity variations described in section 2. Consequently, modelling of the 2018 data has been split into several sections, bounded by the transition between numbered event characteristics, as explained in sections 2 & 3. Estimated velocities for each model run were based on the results of the curve fitting, excluding the region between features 5 and 6, where no clear numbered event was detected.

Figures 13 and 14 show examples of the results of model reproductions of the original observed features in the LOFAR data from 2018 in PL612. Because of the altitude ambiguity discussed previously, there are two versions of each model run, one using the heights and velocities assuming the event occurs in the F-region, and the other the E-region. In each plot the top row shows the original LOFAR data, the middle row shows the model runs for the E-region, and the bottom shows the F-region runs.

Figures 15 and 16 show model runs for numbered events 1-6, and 6-12, in the Cassiopeia A data. Just to note again, that no velocity estimates could be made using curve fitting for the Cassiopeia A data, however a set of models could still be produced by setting the velocity of each section by eye for a fixed altitude. This is due to the fact that the frequency dependence of the intersection between the secondary fringing of one numbered event and the next is seen clearly and the position of this point is approximately only a function of velocity and spacing, when the altitude is held constant. Once more, this information can only be extracted with broadband observations. In all model plots shown, the timescales for the model runs and the original LOFAR data are identical. The modelling also replicates the less defined secondary fringing in Cassiopeia A despite changes in amplitudes and scale sizes of the ionospheric perturbations, due to the difference in angular size of the sources.

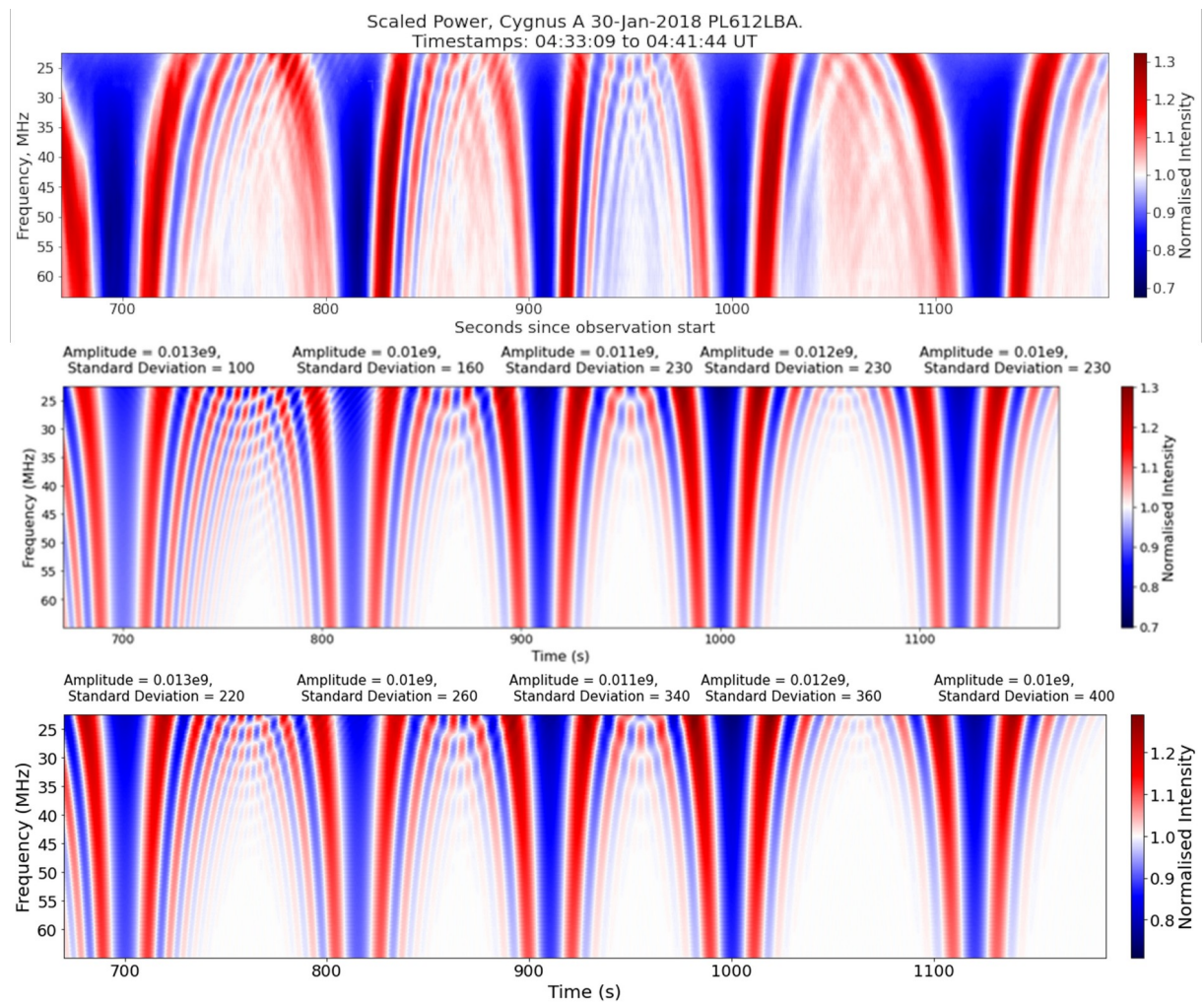

**Figure 13**: Examples of the output model reproductions, here using numbered features 6-10 in the Cygnus A data from PL612 on 30$^{th}$ January 2018. Top row: original LOFAR dynamic spectra for events 6-10. Middle row: Modelled reproduction of features 6-10, assuming E-region propagation with a velocity of 70 ms$^{-1}$. Bottom row: Modelled reproduction of features 6-10, assuming F-region propagation with a velocity of 115 ms$^{-1}$. The values for amplitude and standard deviation define the shape and magnitude of the Gaussian perturbation in each case. The perturbation amplitudes in the model are expressed in rad Hz. A perturbation of $10^{10}$ rad Hz corresponds to approximately 1 TECu of line of sight TEC perturbation

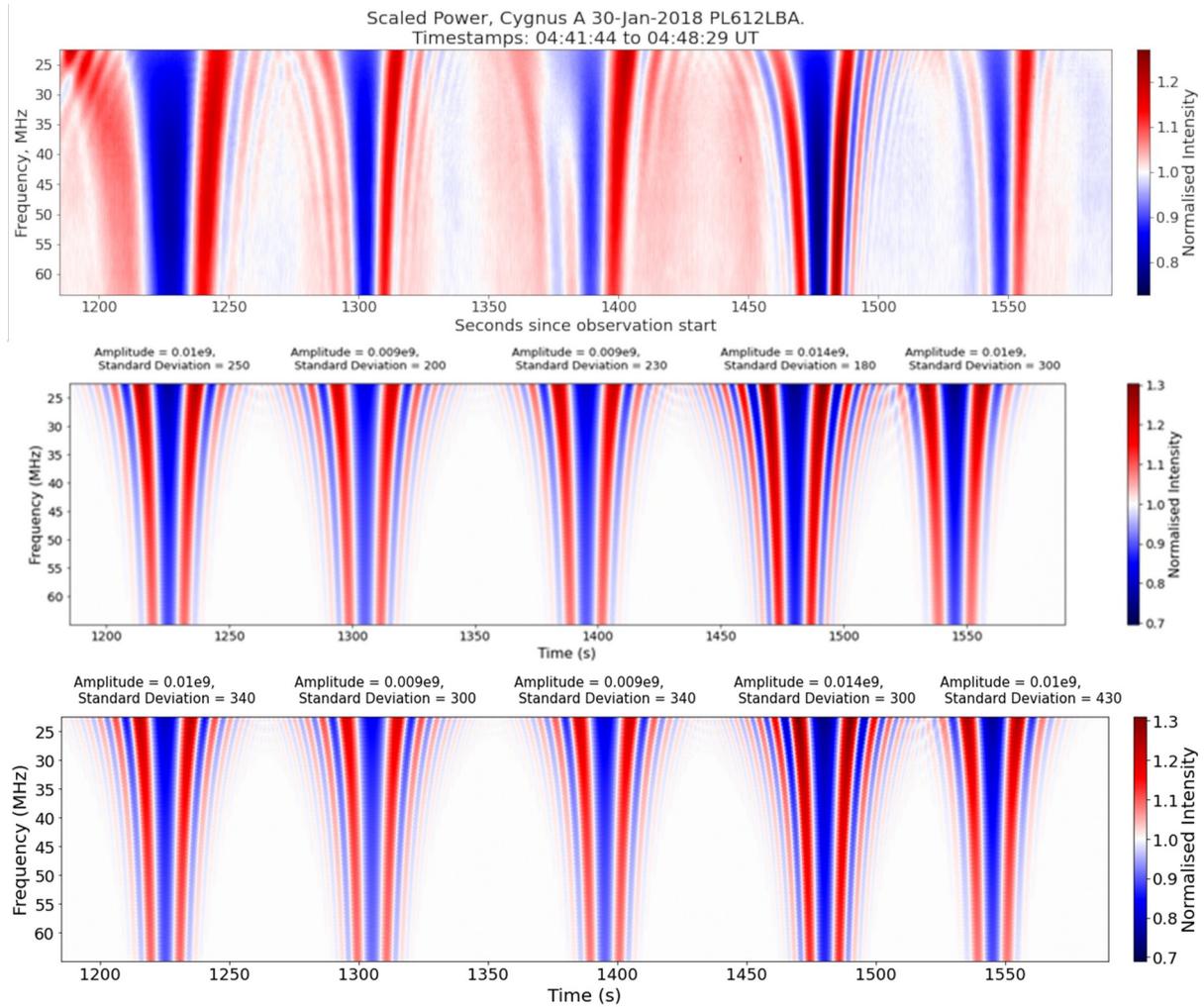

**Figure 14**: Modelled reproductions of features 11-15 in Cygnus A data, from LOFAR station PL612 on 30$^{th}$ January 2018. Top row: Original LOFAR data showing numbered events 11-15. Middle row: Modelled reproductions assuming E-region propagation at a velocity of 120ms$^{-1}$. Bottom row: Modelled reproductions assuming F-region propagation at 200 ms$^{-1}$. Intensity scaling is the same in all plots.

Despite the simplicity of the model, it is able to accurately replicate the general structure and intensity distribution of the features. It can be seen in figure 14 how sensitive the model is to velocity; the first v-shaped signal fade is too narrow compared to the original whereas the fourth one is a little too broad. The model does not accurately replicate some of the asymmetry seen in original signal intensity such as when comparing the intensity and curvature of the left and right side boundary signals (e.g. third feature in figure 14). Asymmetries in the model are constrained by a parameter for horizontal plasma density gradient. However in Boyde et al., (2022), it was demonstrated that arbitrary values for this parameter could be used to replicate observed asymmetries in the LOFAR data, implying that their true physical cause was not horizontal plasma density gradient. It is emphasised that in order to accurately replicate the observed features, modelled TEC perturbations of a few mTECu were necessary, which strongly suggests that the individual v-shaped fades were caused by plasma structures of very small amplitude.

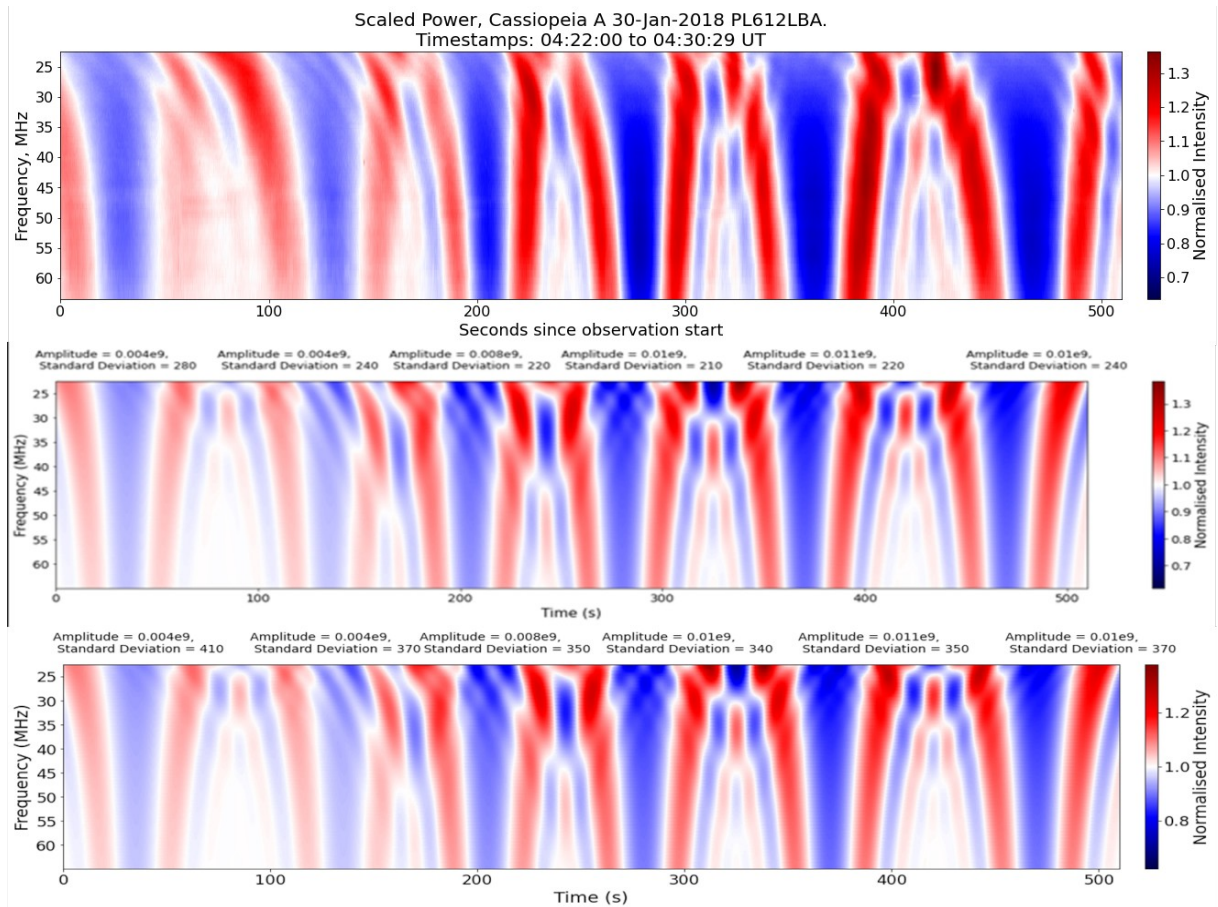

**Figure 15**: Modelled reproductions of features 1-6 in Cassiopeia A data, from LOFAR station PL612 on 30th January 2018. Top row: Original LOFAR data showing numbered events 1-6. Middle row: Modelled reproductions assuming E-region propagation at a velocity of 50ms$^{-1}$. Bottom row: Modelled reproductions assuming F-region propagation at 80 ms$^{-1}$. Velocity estimates are made by aligning the position of the cross-over point between the secondary fringing regions of each event, as this is a function of velocity once altitude is held constant.

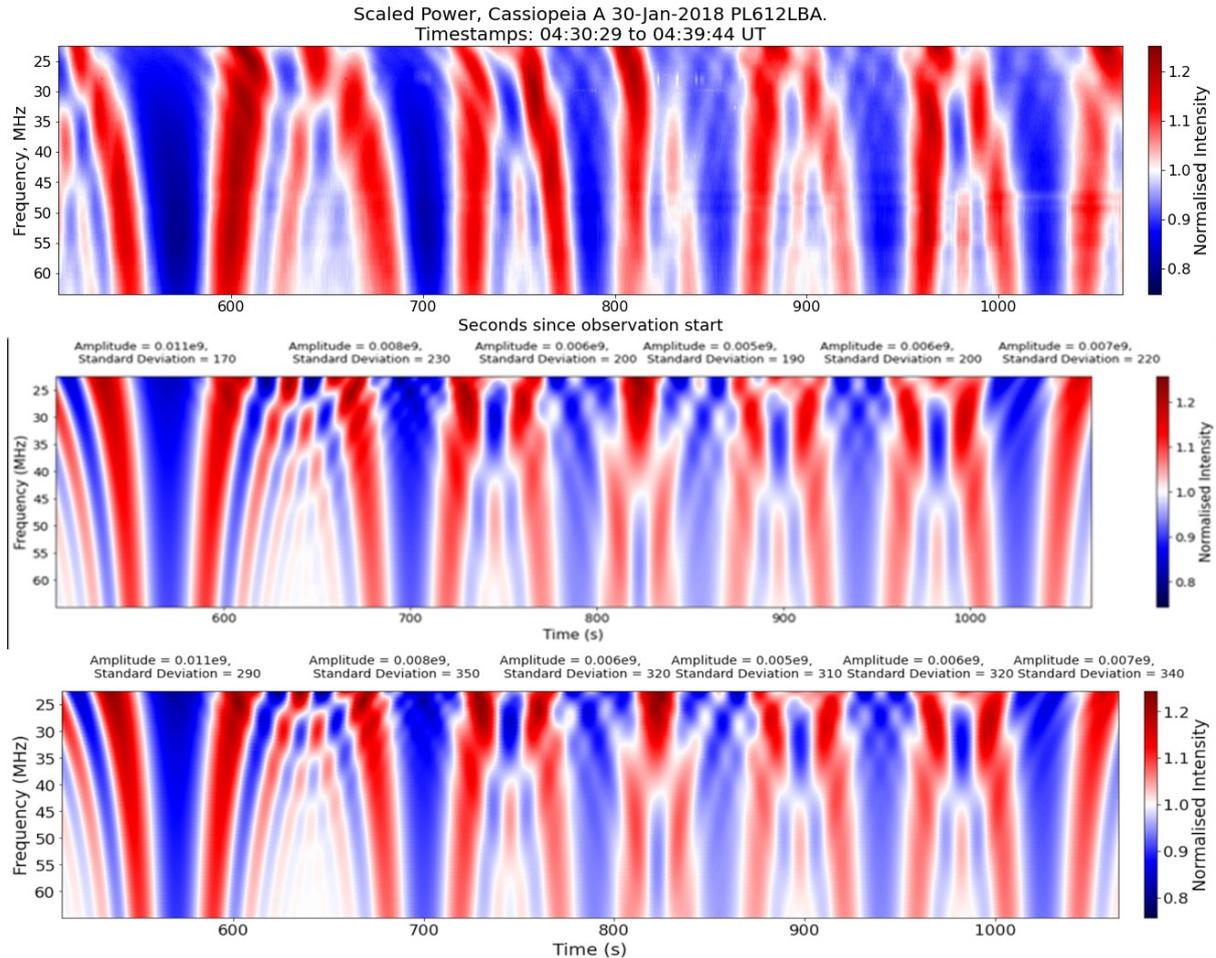

**Figure 16**: Modelled reproductions of features 6-12 in Cassiopeia A data, from LOFAR station PL612 on 30$^{th}$ January 2018. Top row: Original LOFAR data showing numbered events 6-12. Middle row: Modelled reproductions assuming E-region propagation at a velocity of 43 ms$^{-1}$. Bottom row: Modelled reproductions assuming F-region propagation at 70 ms$^{-1}$. Again, velocities are estimated by the positioning of the cross-over point between the secondary fringes of one event and the next.

In Figure 17 the modelled results for the reproductions of the 2016 observations of Cygnus A from UK608 are shown. The top panel shows the original LOFAR data and the middle and bottom panels show E- and F-region modelling respectively. As previously noted, the individual numbered features here are more separated and with extensive secondary fringing regions. Again, the model is well able to reproduce the original LOFAR features; the only differences being subtle asymmetries of the secondary fringing and boundary signal enhancements in a couple of the features. The model also slightly overestimates overall signal intensity; this is possibly a result of an underestimation of sky-noise which affects the original LOFAR signals, particularly at the lower end of the frequency range, but has to be reproduced synthetically in the model. We note that the outputs from the model runs for both 2016 and 2018, while successfully reproducing the original features in the data, are somewhat insensitive to altitude owing to the fact that the scattering characteristics of the features in the original LOFAR data were consistent with a wide range of possible altitudes.

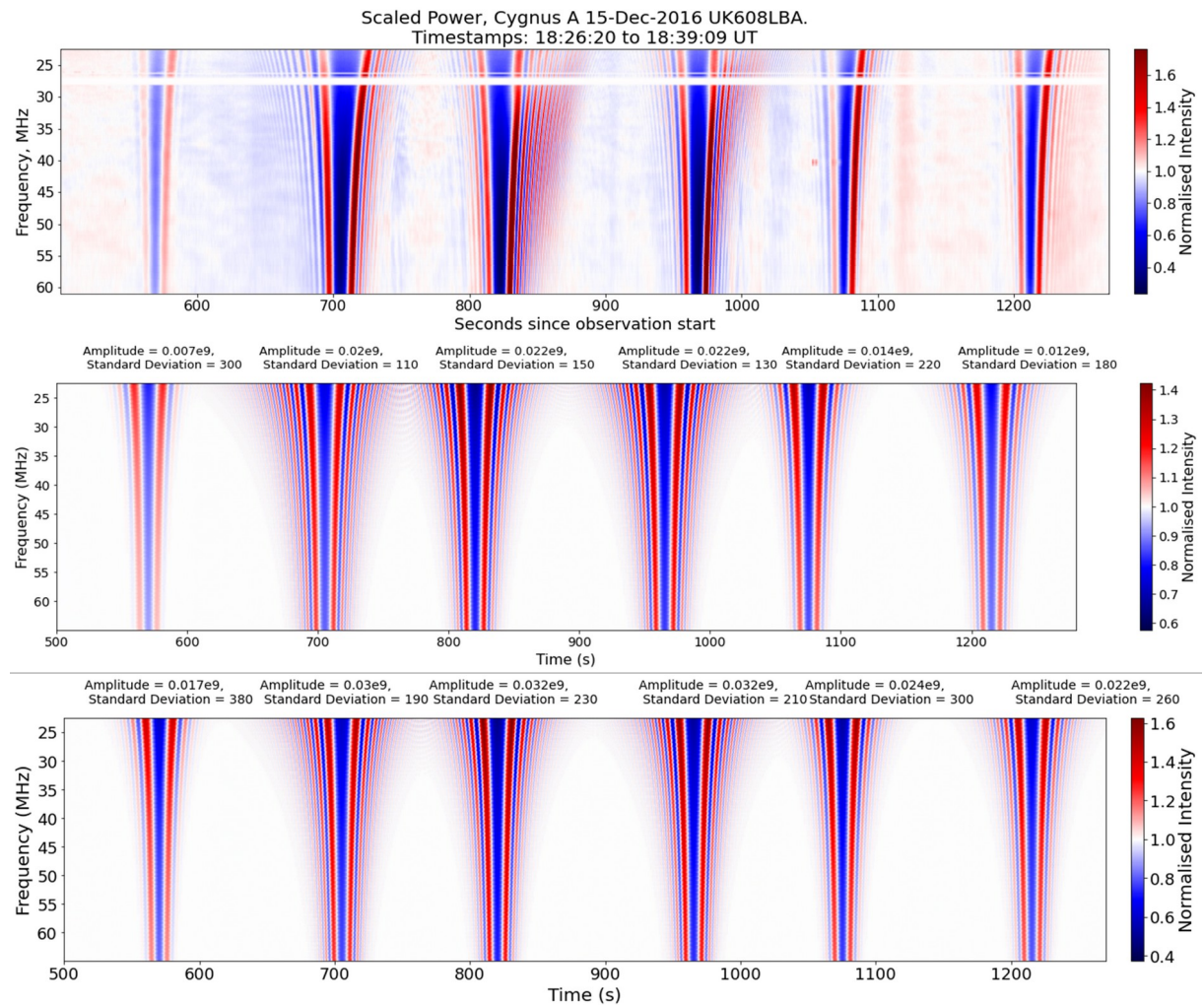

**Figure 17**: Modelled reproductions of the Cygnus A data, from LOFAR station UK608 on 15$^{th}$ December 2016. Top panel: Original LOFAR data showing all numbered events. Middle panel: Modelled reproductions assuming E-region propagation at a velocity of 120 ms$^{-1}$. Bottom panel: Modelled reproductions assuming F-region propagation at 200 ms$^{-1}$. Velocities are estimated from curve fitting. The model runs and the original data are plotted to the same timescales. Model runs for both E- and F-region altitudes are provided here as the LOFAR data alone could not be used to constrain altitude, as reflected in the modelling output.

## 5. Discussion

The form of all the features seen in these data and their successful reproductions are all consistent with type 2 quasi-periodic oscillations in the ionosphere as categorised by Maruyama (1991, 1995). Namely, a distinct signal fade which is symmetrically bounded by large signal enhancements, which we refer to as 'boundary signals', and a series of ringing irregularities with varying degrees of definition depending on observing circumstances.

For the 2018 data we are unable to unambiguously determine the altitude of propagation of the QPOs, and so we have proceeded to analyse this event assuming that it maybe propagating either at F- or E-region altitudes, with altitude values having been acquired from the 2016 IRI model. We find propagation velocities, assuming an E-region altitude, starting at 50 ms$^{-1}$ and accelerating throughout the observation to 120 ms$^{-1}$. If one assumes an F-region propagation then the velocities increase from 80 ms$^{-1}$ at the start to 200 ms$^{-1}$ by the end. Velocity ranges of 50 – 120 ms$^{-1}$, consistent with the QPO propagating in the E-region, is within the typical range of expected velocities for a small to medium scale TID, however the wide range is unusual. This definite change in velocity, regardless of altitude, may be caused by different populations of plasma moving at different altitudes with their own velocities with each dominating the LOFAR signal at different points in the observation. It is noted however that the periodicities of the 2018 data were of order a few minutes which is closer to the Brunt – Väisälä frequency for E-region altitudes than F-region. The large number (>15) of individual signal fades recorded on both radio sources and the fact that this event filled most of the observing window is more consistent with the continuous oscillation type QPOs identified by Yamamoto et al., (1991). It is also noted that Yamamoto et al., (1991) identify continuous oscillations of this kind as being more common in the early morning, which is when the 2018 data were recorded. A continuous oscillation would, in turn, imply some continuous driving process, although determination of the driving processes behind the events observed are outside the scope of this paper.

For the 2016 study, contemporary data was available from two ionosondes located close to the IPP. The appearance of non-blanketing sporadic-E in the ionograms from these data appearing intermittently throughout the observing window, and the absence of any notable disturbance to the F-region traces make a strong case for this example of a QPO being located in the E-region. Here the velocity remained more constant at ~120 ms$^{-1}$, with no strong evidence for acceleration or deceleration. Again, this velocity is reasonably typical of a small to medium scale traveling ionospheric disturbance. However, we could not definitively identify the onset and stop times of this event in most of the LOFAR station which saw it, thus precluding an opportunity to establish in which direction it was propagating. In the IPP maps for this event (figure 5), there are only a few minutes which seemingly separate the appearance of the QPO in LOFAR stations that are quite widely disbursed, geographically speaking. Therefore it is unlikely that each station saw *exactly* the same event but instead that this QPO was part of some larger regional structure which contained many such examples. This event, consisting of 5-6 individual signal fades, is more consistent with a the non-continuous, quasi-periodic oscillation identified by Yamamoto et al., (1991), rather than the continuous type which more accurately defines the 2018 observations.

In the 2018 event, the QPOs were only seen clearly (on both radio sources) from PL612. No other detections were made in other LOFAR stations that were geographically close to

PL612. In the 2016 event, the QPOs were detected in a small number of LOFAR stations in Western Europe. Given that, during such observations, LOFAR lines of sight are parallel and have maximum baselines between stations of ~1000 km, it is highly unlikely that these QPOs were a feature of the solar wind. Scale sizes of the solar wind irregularities which generate interplanetary scintillation are such that all LOFAR lines-of-sight would likely fall within the spatial extent of the same solar wind feature. That the QPOs are strongly localised geographically therefore makes an ionospheric source far more plausible.

The results of the delay-Doppler spectral analysis of both events have produced some of the clearest examples of scintillation arcs in an ionospheric context. Their definition was such that it was possible to utilise them to estimate the propagation velocity of the QPOs. Furthermore, the presence of primary and secondary arcs in the 2018 observations suggest the presence of two scattering populations of plasma in the ionosphere which would be quite consistent with the two-layer E-layer model proposed by Maruyama et al., (2000).

Considerable characterisation and theoretical work on scintillation arcs has been conducted in the closely related field of observations of interstellar scintillation from pulsars (ISS: e.g. Mall et al., 2022; Main et al., 2023). Vast differences in size and timescales aside, another key consideration in ISS is that the scattering medium may occupy a significant proportion of the raypath. In the ionosphere, the scattering region is much thinner, but also much closer to the observer. However, it has been argued that the appearance of parabolic scintillation arcs in ISS reveal an underlying anisotropy to the scattering medium which is not simply stochastic (Walker et al., 2008). A future development of the present work will be to investigate how observations such as those in this study may inform the field of ISS and vice-versa.

Many of the features observed have been successfully reproduced using a Gaussian phase screen model (Boyde et al., 2022). The model accurately reproduces LOFAR signal intensity across the frequencies used and with similarly clear definition of secondary fringing as the original data. The model is also capable of utilising a range of source sizes and shapes; the subtle asymmetries seen here may be more accurately reproduced with some form of skewed Gaussian source. We note that, unlike the two-layer sporadic-E model of QPOs proposed by Maruyama et al., (2000), and despite the appearance of primary and secondary arcs in the DDS for the 2018 study, the simpler single phase screen model used here was sufficient to reliably reproduce the QPOs seen in the dynamic spectra.

Previous observations of QPOs have used either single frequency channel measurements of ionospheric distortions to satellite data, or ground-based VHF radars. To our knowledge these represent the first examples of such features being observed in broadband, with clear frequency dependent behaviour seen. Furthermore, the high temporal and frequency resolution of LOFAR, has allowed us to make very detailed observations of the secondary diffraction fringing regions which accompany the main signal fades.

# Conclusions

Broadband trans-ionospheric radio propagation observations of highly-defined symmetric quasi-periodic oscillations have been made in two case studies, with the International LOFAR Telescope. These oscillations are characterised by a distinct broadband signal fade which is approximately v-shaped when observed in dynamic spectra across the frequency range 22.5-64.8 MHz. The v-shaped fades are bounded on both sides by a series of secondary diffraction fringes. The diffraction fringes are more distinct and with greater spread at the lower frequencies. The first case study, from January 2018, was of a continuous series of such oscillations which persisted in the raypath for the full duration of the 90-minute observing window. The second from data taken in December 2016 was of shorter duration with just 6 v-shaped fades, but with particularly well defined secondary diffraction fringing. These features yielded very clear scintillation arcs in their corresponding delay-Doppler spectra, and were successfully modelled using a Gaussian thin-screen phase model, with amplitudes of < mTECu. They constitute some of the clearest examples of these features thus far reported, and are the first to be observed in broadband.

# Acknowledgments

The LOFAR data for these observations can be accessed from the LOFAR Long Term Archive (LTA: https://lta.lofar.eu), under the project code LC9_001 with observation ID L640829 for the 2018 case study, and under LC7_001 with observation ID L562627 for the 2016 case study. The International LOFAR Telescope is designed and constructed by ASTRON. It has observing, data processing, and data storage facilities in several countries that are owned by various parties (each with their own funding sources) and that are collectively operated by the ILT foundation under a joint scientific policy. The ILT resources have benefited from the following recent major funding sources: CNRS-INSU, Observatoire de Paris and Université d'Orléans, France; BMBF, MIWF-NRW, MPG, Germany; Science Foundation Ireland (SFI), Department of Business, Enterprise and Innovation (DBEI), Ireland; NWO, The Netherlands; The Science and Technology Facilities Council, UK; Ministry of Science and Higher Education, Poland. The ionosonde data were accessed through the Global Ionospheric Radio Observatory, accessible at http://spase.info/SMWG/Observatory/GIRO.

H. Trigg was funded for this work by a student summer bursary grant from the Royal Astronomical Society (PI: G. Dorrian). This work is supported by the Leverhulme Trust under Research Project Grant RPG-2020-140. Ben Boyde acknowledges receipt of a PhD studentship from the same grant. R. A. Fallows was partially supported by the LOFAR4SW project, funded by the European Community's Horizon 2020 Program H2020 INFRADEV-2017-1 under Grant 777442.